\documentclass[11pt,x11names,a4paper]{article}
\pdfoutput=1

\usepackage[utf8]{inputenc}
\usepackage{lmodern}
\usepackage[T1]{fontenc} 
\usepackage{microtype} 

\usepackage[a4paper, left=25mm, right=25mm, top=30mm, bottom=25mm]{geometry} 
\usepackage{fancyhdr} 
\usepackage{abstract}

\usepackage{titlesec} 
\titleformat{\section}[block]{\large\bfseries\centering}{\thesection}{1em}{} 
\titleformat{\subsection}[block]{\bfseries}{\thesubsection}{1em}{} 

\usepackage{xcolor}

\usepackage{cite}
\usepackage{hyperref}
\hypersetup{
	colorlinks=true,
	linkcolor=Blue4,
	citecolor=Red4,
	urlcolor=Green4,
	linktoc=page
}

\usepackage{enumerate}
\usepackage{graphicx}
\usepackage[percent]{overpic}
\usepackage{tikz}
\usepackage{subcaption}
\usepackage[many]{tcolorbox}
\tcbset{highlight math style={enhanced,
  colframe=DeepSkyBlue4,colback=Honeydew1}}

\usepackage{amsmath,amssymb,slashed,mathbbol,mathtools}
\usepackage{amsthm}
\numberwithin{equation}{section}
\usepackage{slashed}

\DeclareSymbolFontAlphabet{\amsmathbb}{AMSb}%

\newcommand{\sech}{\text{sech}\,}

\newcommand{\dd}{\mathrm{d}}

\newcommand{\e}{\mathrm{e}}

\newcommand{\w}{\wedge}

\newcommand{\be}{\begin{equation}}
\newcommand{\ee}{\end{equation}}
\newcommand{\bea}{\begin{eqnarray*}}
\newcommand{\eea}{\end{eqnarray*}}

\newcommand{\f}[2]{\frac{#1}{#2}}
\newcommand{\R}{\mathbf{R}}
\newcommand{\p}[1]{\phantom{#1}}
\newcommand{\vol}{\text{vol}}

\newcommand{\Tr}{\text{Tr}~}
\newcommand{\trace}{\Tr}

\newcommand{\SU}{\mathop{\rm SU}}
\newcommand{\SO}{\mathop{\rm SO}}

\newcommand{\U}{\mathop{\rm {}U}}

\newcommand{\sphere}{\mathbf{S}}
\newcommand{\grSO}{\SO}
\newcommand{\grSU}{\SU}
\newcommand{\grU}{\U}

\newcommand{\dash}{^\prime}
\newcommand{\s}{\sigma}
\newcommand{\dif}{\dd}
\newcommand{\PV}{\text{PV}\,}

\newcommand{\HyperGeo}{{}_2F_1}

\title{\fontsize{18pt}{18pt}\selectfont\textbf{Precision Holography for 5D Super Yang-Mills}\vspace{2mm}}

\author{
\large{
\href{mailto:ffg@hi.is}{Fri{\dh}rik Freyr Gautason} and \href{mailto:vgmp@hi.is}{Valentina Giangreco M. Puletti}}\\[5mm]
{\normalsize University of Iceland, Science Institute}\\
{\normalsize Dunhaga 3, 107 Reykjav{\'i}k, Iceland}
}

\date{}

\begin{document}  
{\hypersetup{urlcolor=black}\maketitle}
\thispagestyle{empty}

\vspace{3cm}
\begin{abstract}
\noindent 
We study 1/2-BPS Wilson loop (WL) operators in maximally supersymmetric Yang-Mills theory (MSYM) on $\sphere^5$. 
Their vacuum expectation value (vev) can be computed exactly at large $N$ thanks to supersymmetric localization. 
The holographic dual to MSYM on $\sphere^5$ is geometrically realized by a stack of $N$ D4-branes with spherical worldvolume in ten dimensions.  We compute the vev of the circular WL using holography by evaluating the partition function of a fundamental string in this background. Our focus is on the next-to-leading order correction to the string partition function which is composed of two parts; the dilaton coupling to the worldsheet and the one-loop fluctuations of the worldsheet itself.
We face a variety of issues, some related to the presence of a non-constant dilaton, and others that are common to its AdS analogue.  However, the universality of UV divergences as well as the importance of a proper choice of an IR regulator have been recently stressed in the literature. Inspired by this, we resolve our issues by first carefully treating the Weyl anomaly which receives contributions from the non-constant dilaton, and then by computing the ratio of our partition function and the one of a string in AdS$_4\times \mathbf{C}P^3$, which is dual to a 1/2-BPS WL in ABJM. Crucially, this approach yields a finite result which matches the corresponding ratio of WL vevs on the gauge theory side. 
\end{abstract}

\newpage

\setcounter{tocdepth}{2}
\tableofcontents

\vskip 1 cm
\noindent

\section{Introduction and Summary}
\label{sec:introduction}

Wilson loops are universal gauge-invariant operators which were originally devised to characterize the vacuum of the given gauge theory \cite{Wilson:1974sk}. 
With the advent of AdS/CFT and localization, supersymmetric extensions of Wilson loops came to the forefront as tools to both study the vacuum structure of gauge theories using holography, and also to further our understanding of holography itself. In this paper we focus on the latter.

The most widely studied holographic Wilson loop (WL) is the circular one in ${\cal N}=4$ supersymmetric Yang-Mills theory (SYM). The standard operator is dressed with  coupling to  adjoint scalars such that the operator preserves 1/2 of the supercharges of the theory. The vast amount of supersymmetry enables the exact computation of the vacuum expectation value (vev) at large rank $N$ using supersymmetric localization \cite{Erickson:2000af,Drukker:2000rr,Pestun:2007rz}. In holography, Wilson loop operators are dual to fundamental strings hanging from the conformal boundary~\cite{Maldacena:1998im}. In particular, for the circular Wilson loop in ${\cal N}=4$ SYM, the dual string is attached to the boundary of AdS$_5$ and extends into the bulk while staying at fixed location on the five-sphere. The partition function of the quantized string should reproduce the Wilson loop vev exactly
\be\label{duality}
\langle {\cal W} \rangle = Z_\text{string}\,.
\ee
In practice, the right hand side can only be evaluated when the string coupling is sufficiently weak and the worldsheet sigma model is weakly coupled. In this case we can utilize a saddle point expansion around a classical string in AdS. The leading order contribution is given by the string action evaluated on-shell which is just its regularized area. The next-to-leading order correction is given by the one-loop string partition function, and so on. This expansion on the string theory side corresponds to the strong coupling expansion on the SYM side, which fortunately we have access to via supersymmetric localization. Finding a precise match between the SYM result and the string theory result beyond leading order has unfortunately proved difficult. Presumably this requires a careful treatment of all measure factors and regularization of UV divergences in the string path integral. A careful treatment was initiated in \cite{Drukker:2000ep} where gauge fixing and ghost determinants were discussed in detail. The one-loop path integral was then computed for the circular string in \cite{Kruczenski:2008zk,Kristjansen:2012nz,Buchbinder:2014nia} using various methods. Collecting all contributions to this order does not lead to a satisfactory match with the field theory result. In fact, even the scaling of the answer with $\lambda$ does not agree. 

The origin of this mismatch has been discussed recently in \cite{Giombi:2020mhz}. 
There it was emphasized that the Wilson loop operator, appearing on the left hand side of \eqref{duality}, should not be normalized with respect to the rank of the gauge group and therefore scales as $\sim N$. 
On the string theory side this effect is reproduced by the dilaton coupling of the string worldsheet which is provided by the so-called Fradkin-Tseytlin (FT) action~ \cite{Fradkin:1984pq,Fradkin:1985ys}. 
The role of the FT action for holographic Wilson loops was previously emphasized in \cite{Lewkowycz:2013laa,Chen-Lin:2017pay}. In AdS$_5$ the FT term means that the tree-level string partition function scales with $g_s^{-1}$ where $g_s ={\lambda\over 4 \pi N}$ is the string coupling constant.\footnote{The coupling to the dilaton on general worldsheet is $g_s^{-\chi}$ where the $\chi$ is the Euler character of the worldsheet.} The FT term therefore affects the $\lambda$ scaling of the string partition function but still does not fully resolve the mismatch when compared to the QFT. Giombi and Tseytlin \cite{Giombi:2020mhz} suggested that the remaining discrepancy should be corrected by a careful treatment of the cancellation of  UV divergences one encounters when computing one-loop string partition functions. The outcome of which should be that the naive string result is multiplied by $(T/2\pi)^{\chi/2}$ where $T$ is the effective tension of the string.%
\footnote{This is the tension felt by the string in a curved background. For a circular string in AdS the classical worldsheet geometry is just AdS$_2$ and the classical action of the string is $S_\text{classical} = -2\pi T$. Since the regularized area of AdS$_2$ is $-2\pi$, the remaining factor in $S_\text{classical}$ can be taken as the definition of $T$ in this case.} 
Some possible explanations for this factor were offered in \cite{Giombi:2020mhz} but  a full clarification of its origin is still lacking.

Another way to deal with the mismatch in the Wilson loop vev prefactor is to compute ratio of Wilson loop vevs. This program was carried out in \cite{Forini:2015bgo,Faraggi:2016ekd,Forini:2017whz,Cagnazzo:2017sny,Medina-Rincon:2018wjs} where the circular WL discussed above was compared to the latitude one. The latitude WL~\cite{Drukker:2007qr,Drukker:2007dw,Drukker:2007yx } can also be evaluated on the QFT side using supersymmetric localization~\cite{Pestun:2009nn,Young:2008ed,Bassetto:2008yf}, but on the string theory side the advantage of computing a ratio of WL vevs is that multiplying factors drop out. 

In this paper we study the circular Wilson loop in five-dimensional SYM on $\sphere^5$ with radius ${\cal R}$. Since SYM is not conformal in $d\ne 4$, placing the theory on a curved manifold including only minimal couplings breaks supersymmetry. In order to preserve all 16 supersymmetries, one must introduce additional couplings in the Lagrangian \cite{Blau:2000xg}. Once this is done, the theory can be localized to a matrix model which enables the computation of free energy and WL vev just as for ${\cal N}=4$ SYM in four dimensions \cite{Kim:2012ava} (see also \cite{Minahan:2015jta,Minahan:2015any,Bobev:2019bvq}). The matrix model turns out to be the same as for pure supersymmetric  Chern-Simons theory in three dimensions. This model has been solved exactly \cite{Marino:2004eq} which enables us to obtain a closed form expression for the WL expectation value at large $N$
\be\label{SYMWL}
\tcbhighmath{
\langle { \cal W} \rangle = \f{N}{\xi}(\e^{\xi}-1) + \mathcal{O}(N^{-1})\,, }
\ee
where $\xi = g_\text{YM}^2 N/(2\pi {\cal R})$ is the dimensionless 't~Hooft coupling of the theory, as discussed in section \ref{sec:QFT}. 

The holographic dual to SYM on $\sphere^5$ was identified in \cite{Bobev:2018ugk} to be a particular analytic continuation and dimensional reduction of AdS$_7\times \sphere^4$ in eleven-dimensional supergravity. We will review this geometry in detail in section \ref{sec:gravity}. A key feature of the gravitational solution is that the dilaton is non-trivial, signalling the non-conformal nature of the theory. Indeed, it is well known that a five-dimensional SYM is naively non-renormalizable but in the UV grows an extra dimension and is UV completed in the six-dimensional (2,0) theory. This is built into the gravitational dual as a dimensional reduction of AdS$_7$. The goal of this paper is to use the holographic dual geometry to compute the vev of the circular WL. In this way we hope to reproduce the first two terms of \eqref{SYMWL} in the large $\xi$ expansion:
\be
\log \langle { \cal W} \rangle = \xi + \log \f{N}{\xi} + \mathcal{O}(\e^{-\xi})\,.
\ee
The first term was previously reproduced by computing the classical string area in \cite{Bobev:2019bvq}, and so here we are mainly interested in the first quantum correction. Along the way we encounter many of the same issues as in AdS$_5$ discussed above, but due to the non-conformal nature of the theory we also have to resolve some new ones.

As we have discussed, the first quantum correction consists of two terms, the one-loop fluctuations of the string worldsheet, and the FT term. 
Since the dilaton is not constant, the FT term gives a non-trivial contribution beyond simply $g_s^{-\chi}$ (see also \cite{Chen-Lin:2017pay}). Moreover, we find that the FT is highly divergent. We believe this to be a direct consequence of the fact that 5D SYM is non-renormalizable, which means that the dilaton grows without bound in the UV. 
We will argue that the divergence of the FT term is cancelled by a similar divergence of the one-loop fluctuation of the string worldsheet. 
We see this indirectly by a Weyl rescaling of the worldsheet metric to the flat one. 
Now the worldsheet Ricci scalar is zero and so the bulk FT term simply vanishes. In fact the surface term also vanishes and so it would seem that we get no contribution from the FT term. This is however naive. As emphasized in \cite{Cagnazzo:2017sny}, the Weyl rescaling is ill-defined at the center of the disk which effectively changes the topology of the worldsheet from a disk to a cylinder. This also means that the Euler characteristic changes from 1 to 0 and the naive evaluation of the FT term gives $g_s^0$. We must therefore add back the FT associated with a small disk at the center of the worldsheet before Weyl rescaling.

In order to compute the one-loop partition function of the string, we adopt the phase shift method \cite{Chen-Lin:2017pay,Cagnazzo:2017sny} utilizing the flat metric. The one-loop partition function is both UV and IR divergent, where the IR divergence is associated with a cutoff radius $R$ close to the center of the worldsheet. The structure of the divergences is $\log Z\sim -\log(\Lambda \e^{-R})$ where $\Lambda$ is a cutoff on the phase shift momentum. As we have discussed, the cancellation of divergences requires a careful understanding of all measure factors and ghost determinants. We will sidestep this problem by instead computing a ratio of string partition functions. As long as the same steps are followed for the computations of two partition functions, we expect the divergences to have exactly the same structure. We verify this explicitly by computing the one-loop partition function for the circular string in AdS$_4\times {\bf C}P^3$ dual to a circular WL in the ABJM theory \cite{Aharony:2008ug} using exactly the same steps as we did for the 5D SYM case.  Our choice to use the ABJM WL is somewhat arbitrary but is preferred since we want to remain within type IIA string theory. Before cancelling the UV and IR cutoffs in a ratio of partition functions, we must translate the IR regulator $R$ to a diffeomorphism invariant cutoff given by the area of the worldsheet which we remove when computing the one-loop determinants~\cite{Cagnazzo:2017sny}.
This translation depends on the string worldsheet metric and is different for the two cases. 
In particular, this introduces a factor reminiscent of the $\sqrt{T/2\pi}$ prefactor proposed by Giombi and Tseytlin \cite{Giombi:2020mhz}.%
\footnote{This is for the worldsheet with a disk topology, we expect that the factor $(T/2\pi)^{\chi/2}$ to be produced in a similar way on higher genus worldsheets.} 
After this is done however, we argue that the UV cutoff $\Lambda$ and IR cutoff $A$ can be cancelled in the ratio of partition functions resulting in a finite answer
\be
\tcbhighmath{\f{Z_\text{SYM}^\text{string}}{Z_\text{ABJM}^\text{string}} = \Big(\f{N_\text{ABJM}}{4\pi\lambda}\e^{\pi\sqrt{2\lambda}}\Big)^{-1}\Big( \f{N_\text{SYM}}{\xi} \e^{\xi}\Big)\,.}
\ee
On the right hand side we see a perfect match with the ratio of vevs of the circular WLs in 5D SYM~\eqref{SYMWL} on the one hand, and ABJM~\eqref{WABJM} on the other \cite{Kapustin:2009kz,Marino:2009jd,Drukker:2008zx}.

We close this summary with a few comments. First, the close connection of 5D SYM and the (2,0) theory in six dimensions implies that our computation should perhaps be rephrased purely in terms of the (2,0) theory. The Wilson loop in SYM corresponds to a BPS surface operator in six dimensions. On the holographic side, instead of computing the string partition function, we should compute the M2 brane partition function in AdS$_7$ with toroidal boundary. We have verified that the classical contribution is identical to the classical string area (see \cite{Mezei:2018url}) but have not attempted to reproduce the quantum correction from a purely eleven-dimensional computation. Recently there has been considerable progress in this direction \cite{Drukker:2020swu,Drukker:2020atp,Wang:2020xkc,Drukker:2020bes}, and it would be interesting to understand whether our result can be rephrased in the M2 surface operator language.

Next, we note that the localization result, reviewed in section  \ref{sec:QFT}, can be used to compute the free energy beyond leading order in the 't  Hooft coupling $\xi$.  It would interesting to reproduce this answer on the gravity side by computing on-shell action using higher derivative corrections to the supergravity action. Naively the $R^4$ correction of eleven-dimensional supergravity should be all that is needed (see for example \cite{Tseytlin:2000sf}), and it is easy to verify that it scales in the correct way. 
However, it is also apparent that the higher derivative corrections are UV divergent when evaluated on-shell and so a careful treatment of the divergences should be carried out to obtain a precise match. We leave this for future work. 

The structure of the remainder of the paper is as follows. In section \ref{sec:QFT}, we review the localization of 5D SYM on $\sphere^5$ and compute the WL vev at large $N$. In section \ref{sec:gravity}, we review the holographic dual geometry in ten dimensions and discuss its relation to AdS$_7\times \sphere^4$ solution of eleven-dimensional supergravity. In section \ref{sec:string} we introduce the fundamental string solution dual to the circular WL and discuss the one-loop action. In section \ref{sec:oneloop} we compute the one-loop partition function using the phase shift method and compare with a similar computation for ABJM in order to find a match with the QFT in section \ref{ratio}. 
We also include three appendices on the details of the one-loop string action  (appendix \ref{app:lagrangian}), the computation of one-loop partition functions of the SYM string (appendix \ref{app:phaseshift}) and of the circular ABJM string (appendix \ref{app:Ads}).

 \section{Super Yang-Mills on $\sphere^5$}
 \label{sec:QFT}
The construction of a maximal supersymmetric gauge theory on the round sphere is non-trivial since introducing only the minimal coupling to the curved metric breaks supersymmetries. Progress in this direction was made in~\cite{Blau:2000xg, Minahan:2015jta}. 
In these works an action for Euclidean maximal supersymmetric Yang-Mills theory (SYM) on a $d$-sphere $\sphere^d$ is obtained by a dimensional reduction from ten-dimensional SYM in flat space, with the introduction of a minimal coupling to the sphere metric and additional interaction terms. On one hand, these break the original flat space $R$-symmetry from $\grSO(1,9-d)$ to $\grSU(1,1) \times \grSO(7-d)$, but on the other hand, guarantee the existence of sixteen real supercharges~\cite{Blau:2000xg, Minahan:2015jta}.
The corresponding Lagrangian~\cite{Blau:2000xg, Minahan:2015jta} is given by 
\be\label{Lagrangian-SYM}
\begin{split}
\mathcal L &= -{1\over 2 g^2_{\rm YM}} \Tr \left( \frac 12 F_{MN} F^{MN}- \bar\Psi \slashed{D} \Psi + {(d-4)\over 2 \mathcal R}\bar \Psi \Gamma^{089} \Psi +{2(d-3)\over \mathcal{R}^2} \phi^A \phi_A+\right.
\\ 
& +\left. {d-2\over \mathcal R^2} \phi_i\phi^i+ {2 i\over 3 \mathcal R} (d-4) \left[\phi^A, \phi^B\right]\phi^C \varepsilon_{A B C}- K_m K^m \right)\,.
\end{split}
\ee
Here, $\mathcal R$ is the radius of the $d$-dimensional sphere where the theory lives on, the indices $M, N=0, \dots, 9$ are the original ten-dimensional  Lorentz indices, which split into the spacetime indices on the sphere $\sphere^d$, and the scalar indices $I, J=0, d+1, \dots, 9$, due to the reduction from the original ten-dimensional SYM theory.  Moreover, the scalar indices $I, J$ are further broken into the scalar indices $i, j=d+1, \dots, 7$, and $A, B=0, 8,9$, due to the terms proportional to the  $\sphere^d$ radius $\mathcal R$ (and its squared) in the above Lagrangian. The ten-dimensional Majorana-Weyl spinors $\Psi$ get reduced to 16 real components obeying the chirality condition $\Gamma_{11} \Psi=\Psi$.  Finally $K_m$ are auxiliary fields. 
The above Lagrangian \eqref{Lagrangian-SYM} is in Lorentzian signature, and it needs to be Wick rotated, which entails the scalar field to transform as $\phi^0 \to i \phi^0$, and the  Lagrangian as $\mathcal L\to - i \mathcal L$.

In this paper, we are interested in the maximal SYM on a five-dimensional sphere $\sphere^5$. 
The $R$-symmetry group is then $\grSU(1,1)\times \grSO(2)$. The theory is Euclidean and so the corresponding space transformation group is $\grSO(6)$. 
The full supergroup of symmetries is the four-dimensional ${\cal N}=2$ superconformal group $\grSU(4\vert 1,1)$. 
It should be noticed that in five dimensions, the coupling constant $g^2_{\rm YM}$ is irrelevant, implying that five-dimensional maximal SYM theories are non-renormalisable. 
At high energies, these theories are UV completed in the six-dimensional $(2,0)$ superconformal field theory \cite{Douglas:2010iu, Lambert:2010iw}. 
For later convenience we introduce the 't~Hooft-like coupling constant%
\footnote{The constant $\xi$ is related to 't Hooft coupling constant $\lambda$ used in \cite{Bobev:2019bvq} by simply $\xi={\lambda\over 2\pi}$.}
\be
\xi = \f{g^2_{\rm YM} N}{2\pi  \mathcal R}\,.
\ee

The theories described by the Lagrangian \eqref{Lagrangian-SYM} can be localized \cite{Pestun:2007rz, Minahan:2015jta, Minahan:2015any, Gorantis:2017vzz}, and the corresponding matrix-model partition function was given (up to instanton corrections) in \cite{Minahan:2015jta, Minahan:2015any, Gorantis:2017vzz} for any $d$.  
The supercharge employed in \cite{Minahan:2015jta} localizes the theory on a locus described by vanishing gauge fields $A_\mu=0$ and scalar fields $\phi_I=0$ for $I\neq 0$, that is with the exception of $\phi_0$. This is the field used to construct a (dimensionless) $N\times N$ Hermitian matrix $M$, after being Wick rotated and rescaled by $\mathcal R$. 
In the large $N$ limit, the gauge fixed partition function can be evaluated in terms of the eigenvalues of the matrix $M$. 
The corresponding large $N$ partition function (neglecting instanton contributions) for the five-dimensional case is
\be\label{Z-matrix-d5}
Z = {1\over N!} \int \prod_{i=1}^N  \dif \mu_i \, e^{-S_\text{eff}}\,,
\ee
where the effective action is given by 
\be\label{Seff-d5}
S_\text{eff}= {2 \pi^2 N \over \xi} \sum_{i=1}^N \mu_i^2- \sum^N_{j\neq i} \sum_{i=1}^N \log \vert \sinh(\pi (\mu_i-\mu_j))\vert\,,
\ee
and $\mu_i$ are the eigenvalues of the $N\times N$ Hermitian matrix $M$. 
In the large $N$ limit the saddle point equation is then 
\be\label{discrete-saddle-d5}
N{2\pi \over\xi}\mu_i= \sum_{j\neq i}\coth\pi(\mu_i-\mu_j)\,, \qquad \qquad i, j=1, \dots, N\,. 
\ee
After introducing an  eigenvalue distribution $\rho$ as follows
\be
\rho(\mu)  = {1\over N} \sum_{i=1}^N \delta(\mu-\mu_i)\,, 
\ee
and taking the large $N$ continuum limit, the saddle point equation \eqref{discrete-saddle-d5} becomes 
\be\label{saddle-eq-all-5d}
{2\pi\over \xi}\mu= \PV \int_{-b}^{b} \rho(\mu\dash)\, \coth (\mu-\mu\dash)\dif \mu\dash\,. 
\ee
The integral equation \eqref{saddle-eq-all-5d} for $\rho$ and $b$ is well known. 
Indeed, the partition function \eqref{Z-matrix-d5} and the consequent equation \eqref{discrete-saddle-d5}  appear in the matrix formulation of Chern-Simons theories on a three-dimensional sphere $\sphere^3$ \cite{Aganagic:2002wv, Marino:2002fk, Tierz:2002jj, Marino:2004eq}.   
The fact that the partition function for maximal SYM on a $\sphere^5$ equals the partition function of Chern-Simons theories on $\sphere^3$ was originally emphasized in~\cite{Kim:2012ava}.
The solution to the integral equation \eqref{saddle-eq-all-5d} is given by \cite{Marino:2004eq}
\be\label{rho-exact-5}
\rho(\mu)={2 \over \xi} \arctan\left({\sqrt{e^{\xi}-\cosh^2\left({\pi\mu}\right)} \over \cosh\left({\pi \mu}\right)}\right)\,,
\ee
and 
\be\label{b-exact-5}
b={1\over \pi}{\rm arccosh}\,(e^{\xi/2})\,.
\ee
Notice that the solution \eqref{rho-exact-5}-\eqref{b-exact-5} is {\it exact} in the 't~Hooft coupling $\xi$. At leading order in the strong coupling expansion, the eigenvalue density $\rho$ and the extreme of integration $b$ reduce to
\be
\lim_{\xi\to\infty} \rho(\mu) ={\pi\over \xi} \,, \qquad \lim_{\xi\to\infty} b= {\xi\over 2\pi}\,, 
\ee
in perfect agreement with the leading-order results obtained in \cite{Bobev:2019bvq}. 

Before discussing the Wilson loop, we report the large $N$ result for the free energy for the maximal SYM on $\sphere^5$, 
\be
{F \over N^2}={2 \pi^2\over \xi} \int_{-b}^b \rho(\mu) \mu^2 \dif \mu- \int_{-b}^b \dif \mu \, \rho(\mu) \int_{-b}^b \dif \mu\dash \rho(\mu\dash) \log\vert \sinh(\pi (\mu-\mu\dash))\vert\,,
\ee
which can be computed from the effective action \eqref{Seff-d5} in the continuum limit. The planar result can be read from the Chern-Simons free energy on $\sphere^3$~\cite{Marino:2004eq}
\be
{F \over N^2}= -{\xi\over 6}+{\pi^2\over 6\xi } -{\zeta(3)\over \xi^2}+ \mathcal O(e^{-\xi})+ \mathcal O\left({1\over N}\right)\,.
\ee
The leading order was also obtained in \cite{Kim:2012ava, Kallen:2012zn, Minahan:2013jwa, Bobev:2019bvq} from a five-dimensional point of view. 

\subsection{$\frac 12$-BPS Wilson loop expectation value from localization}
In this paper our main focus is on the $\frac 12 $-BPS Wilson loop operator and its vacuum expectation value. As discussed in \cite{Bobev:2019bvq} its vev can be computed using the localization procedure sketched above. Here we extend the result of \cite{Bobev:2019bvq} to all orders in the coupling constant $\xi$ but remaining at large $N$. 
The  Wilson loop in question wraps the equator of the 5-sphere and its expectation value is 
\be
\langle \mathcal W \rangle  = \left\langle {\rm Tr} \left(P e^{i \oint A_\mu \dif x^\mu+ i \oint \dif s \, \phi^0} \right)\right\rangle\,,
\ee
where $A_\mu$ is the five-dimensional gauge fields and $\phi^0$ is the ``timelike'' scalar that does not vanish on the localization locus. 
The gauge field on the other hand does vanish and the WL vev can be evaluated by taking the continuum limit and keeping only  leading term in the large $N$ expansion (see also~\cite{Kim:2012qf})%
\footnote{We have restored an explicit factor of $N$ which was omitted in \cite{Bobev:2019bvq}  due to a different normalization convention for the WL operator.}
\be\label{W5-final}
\langle  \mathcal W \rangle  = N \int_{-b}^b \rho(\mu) e^{2\pi \mu} \dif \mu+ \mathcal O\left({1\over N}\right)=\frac{N}{\xi }\left(e^{\xi}- 1 \right) +\mathcal O\left({1\over N}\right)\,.
\ee
This is the expectation value of a $\frac 12$-BPS Wilson loop located on the equator of the sphere $\sphere^5$ at large $N$ but for {\it any} 't~Hooft coupling $\xi$.%
\footnote{We refer the reader to \cite{Gopakumar:1998ki} for analogous results for Wilson loops in  Chern-Simons theories on $\sphere^3$.}
For large values of the 't Hooft coupling constant $\xi$, the $\frac 12$-BPS Wilson loop expectation value approaches to 
\be\label{logW5-exp-cl}
\lim_{\xi \to \infty} \log \langle  \mathcal W \rangle =\xi \,, 
\ee
which is the classical result derived in \cite{Bobev:2019bvq} both from field theory and supergravity. 
Considering the next-to-leading order in $\xi$, we have for the $\frac 12$-BPS Wilson loop VEV
\be\label{WL-strong-coupling-exp}
\langle  \mathcal W \,\rangle  = N {e^\xi\over \xi}+\mathcal O(e^{-\xi})+ \mathcal O\left({1\over N}\right)\,.
\ee
The goal of the next sections is to reproduce the $\frac 12$-BPS Wilson loop VEV in a string theory setting. In particular, 
the exponential behaviour in the large $\xi$-expansion corresponds to (minus) the classical action of the dual string \cite{Bobev:2019bvq},  cf. section \ref{classical}, while the prefactor is encoded in the one-loop string partition function, cf. sections \ref{sec:oneloop}-\ref{ratio}.

\section{Spherical D4 branes}
\label{sec:gravity}

\subsection{Solution of type IIA$^*$}
The holographic dual to SYM on $\sphere^5$ was constructed in \cite{Bobev:2018ugk} by first solving BPS equations in seven-dimensional maximal supergravity, and then subsequently uplifting to ten dimensions. Since we use slightly different coordinates here, we will review the full supergravity solution. 
Given that we are working with Euclidean branes, the proper framework are the so-called type II$^*$ theories of Hull\cite{Hull:1998ym,Hull:1998vg,Hull:1998fh}. The only difference with the standard type II theories is that the RR-fields are purely imaginary.\footnote{The role of II$^*$ theories will not play a fundamental role in this paper and the imaginary form fields can be thought of as a result of a formal analytic continuation of the supergravity background.}
The ten-dimensional metric is given by
\be\label{metricsigma}
\dd s_{10}^2 = \ell_s^2(N\pi\e^{\Phi})^{2/3}\,\Bigg[\f{4\big(\dd\sigma^2+ \dd\Omega_5^2\big)}{\sinh^2\sigma}+ \dd\theta^2 + \cos^2\theta\, \dd s_{\text{dS}_2}^2  + \f{\sin^2\theta\, \dd \phi^2}{1-\f{h^2}{4}\tanh^2\sigma\, \sin^2\theta} \Bigg]\,,
\ee
where the ten-dimensional dilaton $\Phi$ is 
\be\label{dilaton-10}
\e^{\Phi} = \f{\xi^{3/2}}{N\pi}\bigg(\coth^2\sigma-\f{h^2}{4} \sin^2\theta\bigg)^{3/4}\,.
\ee
This background exhibits
\be
\SO(6)\times \SO(1,2)\times \U(1)\,
\ee
continuous symmetry in complete agreement with the field theory. The five-sphere $\dd\Omega_5^2$ is where the field theory is living and $0\le\sigma<\infty$ plays the role of a radial direction. The solution depends on three (dimensionless) parameters $\xi$, $h$, and $N$. 
The integer $N$ denotes the number of D4-branes and is taken to be large to ensure that the length scales set by the metric is large in string units. 
Next, we have $\xi$ which we already encountered in section \ref{sec:QFT} and is related to the Yang-Mills coupling constant in the QFT. 
Finally, we seem to have one more parameter $h$. For all $0\le h \le 2$ the metric defines a regular background of type IIA$^*$ supergravity. Note that for $h=0$ the symmetry of the background is enhanced to $\SO(5)\times \SO(1,4)$ which does not match the expected symmetry of the QFT. Indeed, in \cite{Bobev:2018ugk} it was determined that for $h=1$ we obtain the relevant supersymmetric background dual to SYM on $\sphere^5$. In this paper we will focus on $h=1$, but will keep $h$ unfixed throughout the computation.

In addition to the metric and dilaton we have the gauge potentials
\be\label{BC-fields}
\begin{split}
B_2 &= \f{h \xi\ell_s^2}{2}\cos^3\theta\,\vol_{\text{dS}_2}\,,\\
C_1 &= \f{hi N\pi\xi\ell_s}{2}(N\pi\e^{\Phi})^{-4/3}\sin^2\theta\,\dd\phi\,,\\
C_3 &= -iN\pi\ell_s^3\cos^3\theta\,\dd\phi\wedge\vol_{\text{dS}_2}\,.
\end{split}\ee
From these we can compute the NSNS and RR field strengths
\be\label{BC-fields-v2}
H_3=\dd B_2\,,\quad F_2 = \dd C_1\,,\quad F_4 = \dd C_3 - H_3 \w C_1\,.
\ee

In \cite{Bobev:2019bvq}, the geometry \eqref{metricsigma} was used to compute the holographic free energy of SYM on $\sphere^5$. This was done by first reducing to six-dimensional supergravity, and then evaluating the regularized on-shell action. In addition to standard infinite counterterms required to regularize the bare evaluation of the on-shell action, a number of finite counterterms had to be considered. These ultimately allowed for a successful match with the QFT.

As mentioned above, the metric \eqref{metricsigma} is completely regular. In particular, for large $\sigma$ the metric takes the form
\be
\dd s_{10}^2 \to \ell_s^2(N\pi\e^{\Phi})^{2/3}\,\Bigg[16\big(\dd r^2+ r^2\dd\Omega_5^2\big)+ \dd\theta^2 + \cos^2\theta\, \dd s_{\text{dS}_2}^2  + \f{\sin^2\theta\, \dd \phi^2}{1-\f{h^2}{4} \sin^2\theta} \Bigg]\,,
\ee
where we have changed coordinates $r=\e^{-\sigma}$. Here we see that as $r\to 0$, the five-sphere smoothly shrinks down to zero size without introducing irregularities in the metric (or other supergravity fields). In the opposite limit, as $\sigma\to0$ we get back the flat-space D4 brane solution:
\be\label{metricUVlimit}
\dd s_{10}^2 \to \xi \ell_s^2\left[4U^{3/2}  \dd\Omega_5^2 + \f{\dd U^2 + U^2 \dd s_{\text{dS}_4}^2 }{U^{3/2}}\right]\,,
\ee
where we have momentarily changed coordinates $\sinh\sigma = U^{-1/2}$ and we have combined the metric on dS$_2$ with the $\theta$ and $\phi$ coordinates to form the metric on dS$_4$. The coordinate $U$ measures the distance to the stack of branes which are formally located at $U=0$. 
Here, we clearly see that the five-sphere acts as the brane world-volume, whereas the remaining five directions are transverse to the brane. We also note that the isometry of the solution is enhanced in this limit to $\SO(1,4)$ which is the R-symmetry of the UV field theory.

\subsection{11D solution}
The metric for D4 branes in flat space develops a singularity in the far UV ($\sigma\to0$ or $U\gg1$) where also the dilaton blows up
\be
\e^{\Phi} = \f{\xi^{3/2}U^{3/4}}{N\pi}\,.
\ee
This solutions should therefore be reinterpreted in eleven dimensions as the geometry around a stack of M5 branes. As explained in \cite{Hull:1998ym,Hull:1998vg,Hull:1998fh,Bobev:2018ugk}, the solutions of type IIA$^*$-theory  are uplifted to the exotic M$^*$-theory with  $(2,9)$ signature. This means that the extra coordinate introduced during the uplift is in fact timelike.

For completeness we review here the uplift of the spherical D4 brane solution in \eqref{metricsigma}.  The eleven-dimensional metric is obtained by combining the ten-dimensional metric with the dilaton and $C_1$ form (see for example \cite{Bobev:2018ugk}), 
\be\label{11DMetric}
\dd s_{11}^2 = L_{\text{AdS}_7}^2\,\Bigg[\f{\dd\sigma^2+ \dd\Omega_5^2}{\sinh^2\sigma}-\f{\dd t^2}{\tanh^2\sigma}+ \f14\Big(\dd\theta^2 + \cos^2\theta\, \dd s_{\text{dS}_2}^2  + \sin^2\theta\, (\dd \phi-h\dd t)^2 \Big) \Bigg]
\ee
where the eleventh coordinate is $x_{11}=2N\pi i\ell_s t/\xi$ and is taken to be imaginary to implement the timelike uplift. We note that $t$ is periodic with periodicity $\xi/N$, and the AdS$_7$ length scale $L_{\text{AdS}_7}$ is related to ten-dimensional quantities through
\be\label{LAdS7}
L_{\text{AdS}_7}^3 = 8\pi N \ell_s^3\,.
\ee
The three-form is constructed similarly yielding
\be
A_3 = -\f{i L_{\text{AdS}_7}^3}{8}\cos^3\theta\,\left(\dd\phi-h\dd t\right)\wedge\vol_{\text{dS}_2}\,.
\ee
We note that the parameter $h$ can be absorbed into a coordinate redefinition $\phi\mapsto \tilde \phi + h t$. 
In fact, this is just the metric on AdS$_7\times$dS$_4$ which is the near horizon geometry of $N$ M5 branes in the M$^*$-theory. This geometry is the holographic dual to the six-dimensional $(2,0)$ theory with non-compact R-symmetry.

\section{Holographic Wilson loop}
\label{sec:string}

The holographic dictionary instructs that the vev of a supersymmetric Wilson loop can be computed by evaluating the partition function of a string which satisfies the boundary conditions compatible with the WL~\cite{Maldacena:1998im}. To leading order in the large $\xi$ limit, the partition function reduces to the on-shell action of the string. In this paper we are interested in the subleading correction to this leading order answer, and so we expand the partition function to second order in the coupling $\xi$:
\be
\label{log-Z-tocompute}
\log Z \approx -S_\text{classical}-S_\text{FT} + \log\text{Sdet}^{-1/2} \mathbb{K}= -S_\text{classical}-S_\text{FT}-\Gamma_{\mathbb{K}}\,.
\ee
In this expression the partition function $Z$ of the string is expanded in terms of the classical action $S_\text{classical}$ at order $\xi^1$ as well as two ``quantum'' correction at order $\xi^0$. The first correction term is the Fradkin-Tseytlin action $S_{\rm FT}$ evaluated on-shell, which we review below. The second correction, $\Gamma_{\mathbb{K}}$, is the one-loop partition function of  bosonic and fermionic fluctuations around the classical configuration of the string. Since only the leading order action for the fluctuations is kept, the path integral is Gaussian and reduces to the determinant of bosonic and fermionic operators. We collectively denote the second order operators by $\mathbb{K}$, and their  determinants by $\text{Sdet} \mathbb{K}$.

Before embarking on the journey of computing these three terms, we give a general discussion of the ingredients in the string action and introduce our notation.
The worldsheet action of the string we will use, consists of three parts
\be
S = S_\text{bosons} + S_\text{fermions} + S_\text{FT}\,.
\ee 
First we have the Polyakov action%
\footnote{Our worldsheet is Euclidean which explains the $i$ multiplying the two-form.}
\be\label{BosonicAction}
S_\text{bosons} = \f{1}{4\pi\ell_s^2}\int\Big( \gamma^{ij}G_{ij}\vol_\gamma  + 2 i B_{ij}\dd x^i \w \dd x^j\Big)\,,
\ee
where $i,j=1,2$ denotes the two-dimensional worldsheet indices and $\gamma_{ij}$ is the worldsheet metric. Here $G_{\mu\nu}$ are the ten-dimensional metric components and $B_{\mu\nu}$ are the components of the 2-form field $B_2$. The notation $G_{ij}$  where we use two-dimensional indices on a ten-dimensional object refers to the pull-back of the ten-dimensional tensor down to two dimensions
\be
G_{ij} = G_{\mu\nu}\partial_i X^\mu \partial_j X^\nu\,,
\ee
where $X^\mu$ are the ten scalar fields living on the worldsheet, and in this context, can be thought of as defining the embedding of the string into the ten-dimensional geometry.

Next, we have the so-called Fradkin-Tseytlin (FT) \cite{Fradkin:1984pq,Fradkin:1985ys} action which introduces a dilaton coupling on the worldsheet
\be\label{FTAction}
S_\text{FT}=\f1{4\pi}\int_M \Phi R_\gamma\vol_\gamma+\f1{2\pi} \int_{\partial M} \Phi K\dd s\,,
\ee
where $K$ is the geodesic curvature and $\dd s$ is the reparametrization invariant measure on the boundary. 
We have included the boundary term to ensure that the dilaton coupling correctly accounts for string loop counting even on worldsheets with boundaries. For constant dilaton this simply gives $S_\text{FT} = \chi \Phi_0$ as expected. 

Two remarks are in order about the FT action which will be crucial below. First, we note that in the large $\xi$ expansion,%
\footnote{Or in fact any derivative expansion, where the momenta of the string modes are small compared to the length scales set by the classical geometry.} 
the FT action \eqref{FTAction} should be thought of as subleading when compared with the bosonic action \eqref{BosonicAction} above. This can be seen from the fact that the bosonic action \eqref{BosonicAction}  is of order $T=1/2\pi\ell_s^{2}$, whereas the FT term  \eqref{FTAction} is of order $T^0$. 
The second related point is that the FT action  \eqref{FTAction} classically violates Weyl invariance of the worldsheet theory: the classical energy-momentum tensor computed from the FT action has non-zero trace. However, as we will review below, the classical Weyl ``anomaly'' of the FT action is exactly compensated by the one-loop quantum Weyl anomaly of the bosonic theory in \eqref{BosonicAction} as well as the fermionic terms discussed below. This pattern of cancellation of Weyl anomaly is expected to carry on to subleading orders such that for example, the one-loop Weyl anomaly of the FT term cancels the two-loop anomaly of $S_\text{bosons}+S_\text{fermions}$ and so on.

Finally, the Green-Schwarz (GS) action which couples the ten-dimensional, 32 component worldsheet GS fermions $\theta$ to the background geometry reads \cite{Cvetic:1999zs}
\be
\label{FermionicAction}
S_\text{fermions} = -\frac{1}{2\pi \ell_s^2}\int\Big\{ i\bar\theta P^{ij} \Gamma_i D_j \theta - \frac{i}{8}\bar\theta P^{ij} \Gamma_{11}\Gamma_i^{\p{i}\mu\nu}H_{j\mu\nu} \theta+\frac{i}{8} \e^{\Phi}\bar\theta P^{ij}\Gamma_i(-\Gamma_{11}\slashed{F}_2+\slashed{F}_4)\Gamma_j\theta\Big\}\,,
\ee
where $\Gamma_\mu$ are the ten-dimensional gamma matrices,
$\Gamma_{11}$ is the chirality operator and 
\begin{equation}
P^{ij} = \sqrt{\gamma} \gamma^{ij} - i\epsilon^{ij}\Gamma_{11}\,.
\end{equation}
Once again, the pull-back of ten-dimensional indices is implied in our notation, for example $\Gamma_i = \partial_i X^\mu \Gamma_\mu$.
In this paper we will work in static gauge where the two worldsheet coordinates are directly identified with corresponding ten-dimensional coordinates. This means that $\partial_i X^\mu = \delta_i^\mu$. 

\subsection{The string configuration}

We consider a Wilson loop wrapping the equator of $\sphere^5$. This is dual to a fundamental string wrapping the same $\sphere^5$ and extending along the $\sigma$-coordinate. 

The classical string solution was presented in~\cite{Bobev:2019bvq}.  In static gauge, the two worldsheet coordinates are identified with the ten-dimensional coordinates $\sigma$ and $\tau$ and $G_{ij} = \gamma_{ij}$. 
Here the coordinate $\tau$ has been introduced to parametrize the equator of $\sphere^5$. Minimizing the action leads to the remaining scalars $X$ to be constant at
\be
\label{classicalSol}
 \text{equator of $\sphere^5$}~,\qquad  \theta = 0~,\qquad \text{any fixed point on dS$_2$}~.
\ee
The worldsheet metric is then
\be\label{classicalMetric}
\dd s_2^2= \e^{2\rho}\Big(\dd\sigma^2+ \dd\tau^2\Big)~,
\ee
where 
\be\label{D4metric}
\e^{2\rho} = \f{4\xi \ell_s^2}{\tanh\sigma\sinh^2\sigma} \,,
\ee
is the conformal factor. In the conformal coordinates used in \eqref{classicalMetric}, the volume form is given by $\vol_\gamma = \e^{2\rho}\dd\sigma\w\dd\tau$ and the curvature scalar is 
\be\label{ricciscalar}
\e^{2\rho}\,R_\gamma = -2\partial_\sigma^2\rho=\f{\sech^2\sigma-4}{\sinh^2\sigma}\,.
\ee

The classical string solution can be uplifted to an M2-brane  in eleven dimensions. The M2 brane wraps the eleven dimensional directions $\tau$ and $t$ (see \eqref{11DMetric}), and extends in the $\sigma$ direction. The metric on the M2-brane worldvolume is
\be\label{M2-metric}
\dd s_{\text{M2}}^2 = L_{\text{AdS}_7}^2\,\Bigg[\f{\dd\sigma^2+ \dd\tau^2}{\sinh^2\sigma}-\f{\dd t^2}{\tanh^2\sigma}\Bigg]\,,
\ee
which is just the metric on AdS$_3$ with boundary  topology  ${\bf T}^2$.

\subsection{The classical on-shell action}
\label{classical}

In this section we review the classical action of the string described by the embedding \eqref{classicalSol}. 
As mentioned in section \ref{sec:introduction}, this was discussed in~\cite{Bobev:2019bvq}, however here we provide a different way to regularize the classical string action by its Legendre transform~\cite{Drukker:1999zq}. 

First, we notice that the pull-back of the $B$-field \eqref{BC-fields} vanishes. 
Hence, the on-shell action of the string \eqref{BosonicAction} takes the simple form
\be\label{Scl}
S_\text{classical} = \f{1}{2\pi\ell_s^2}\int \e^{2\rho} \,\dd \sigma \,\dd \tau = 4\xi\int \f{\dd \sigma}{\tanh\sigma\sinh^2\sigma} = \f{2\xi}{\varepsilon^2} - \f{2\xi}{3} + \mathcal{O}(\varepsilon)~.
\ee
This diverges at the boundary $\sigma=\varepsilon\to 0$. In order to regularize the integral we must Legendre transform to the variables that are conjugate to the transverse directions to the Wilson loop \cite{Drukker:1999zq} (see also \cite{Agarwal:2009up,Young:2011aa} for a similar setup to ours). These are properly identified in the UV limit of our ten-dimensional solution. 
This limit was already discussed in section  \ref{sec:gravity} and the metric was given in \eqref{metricUVlimit}. The appropriate Legendre transformation should be done with respect to the UV coordinates, i.e. the five angles parametrizing $\dd \Omega_5^2$, and the five flat space coordinates used to parametrize 
\be
\dd s_{\mathbf{R}^{1,4}}^2=\dd U^2 + U^2 \dd s_{\text{dS}_4}^2\,, \qquad \text{where} \qquad U=\sinh^{-2}\sigma\,.
\ee 
The latter we can write as $x^a = U \hat\theta^a$ where $a=1,\cdots,5$. The angular variables $\hat\theta^a$ parametrize a unit radius dS spacetime through the constraint $\hat\theta^a\hat\theta^b\eta_{ab} = 1$ and $\eta_{ab}=\text{diag}(1,1,1,1,-1)$. Only these five flat space coordinates are of interest to us as they explicitly depend on $U=\sinh^{-2}\sigma$. We now compute the term required for the Legendre transform and treat that as a counterterm for the classical action\footnote{We ignore the $B$-field term as it does not play a role for our solution.}
\be
S_\text{ct}=-\int\partial_i\left(X^\mu \f{\delta {\cal L}}{\delta \partial_i X^\mu}\right) \dd\sigma\dd\tau= -\f{1}{2\pi\ell_s^2}\int\sqrt{\gamma}\gamma^{ij}\partial_i(X^\mu\partial_j X_\mu)\dd\sigma\dd\tau\,.
\ee
Reexpressing this as a boundary integral at fixed small $\sigma=\varepsilon$, and using the flat $D$-brane coordinates just defined, we find
\be\label{Sct-0}
S_\text{ct}=\f{\xi}{2\pi} \int_{\partial M}\f{\partial_\sigma U}{U^{1/2}} \dd \tau= -\f{2\xi}{\varepsilon^2}-\f{\xi}{3}+ \mathcal{O}(\varepsilon)\,.
\ee
Combining the bulk term \eqref{Scl} with the counterterm \eqref{Sct-0}, we obtain the final answer
\be
\label{s-classical}
S_\text{classical} + S_\text{ct}= -\xi\,.
\ee
This result can also be obtained by evaluating the classical volume of the M2 brane \eqref{M2-metric}
\be
S_{\text{M2}} = \f{1}{(2\pi)^2\ell_s^3} \int  \sqrt{-g_{\text{M2}}}\, \dd\sigma\,\dd\tau\,\dd t\,.
\ee
Using the periodicity of $t$ and the relation \eqref{LAdS7} we recover \eqref{Scl}. In this case the volume can be regularized by introducing a boundary counterterm which is proportional to the boundary area of the M2 brane and the final result is again \eqref{s-classical}.

When compared with the minus logarithm of the Wilson loop expectation value \eqref{logW5-exp-cl} as computed in the QFT in section \ref{sec:QFT}, we see that this precisely agrees with the leading order answer in the large $\xi$ expansion. 
For the reminder of this paper we will focus on extracting the next-to-leading order contribution in string theory and compare it with the QFT result.

\subsection{The one-loop string action}

In this subsection we work out the string action at order $\xi^0$, which comprises two terms, the FT action and the action for the quantum fluctuations. 
As we explained above, the FT action gives a contribution which is of order $\xi^0$ even though it is a classical term in the string action.

\subsubsection{Fradkin-Tseytlin action}

In order to  evaluate the FT action \eqref{FTAction} on-shell, we need the pull-back of the dilaton \eqref{dilaton-10}, which is
\be\label{dilaton}
\e^{2\Phi_0}\equiv P[\e^{2\Phi}]=\f{\xi^3 }{ N^2 \pi^2}\coth^3\sigma \,.
\ee
This should be evaluated directly in the action \eqref{FTAction}. For this we need in addition to the curvature scalar in \eqref{ricciscalar}, the geodesic curvature on the boundary which is located at a fixed small $\sigma$. This is easily computed in conformal coordinates, and takes the form
\be
K \dd s = (\nabla^\mu n_\mu)\e^\rho\dd\tau=\partial_\sigma\rho\,\dd\tau\,.
\ee
Combining these expressions, we obtain
\be\label{FTtermdivergent}
S_\text{FT}=-\int_\varepsilon^\infty\Phi_0\partial_\sigma^2\rho\,\dd\sigma + \Phi_0\partial_\sigma \rho\Big|_{\sigma = \varepsilon} = \f{9}{4\varepsilon} + {\cal O}(\varepsilon^0)\,,
\ee
where we have performed the integration over the angular variable $\tau$. 
We easily see that even including the boundary term, this expression has divergences which do not cancel in the limit $\varepsilon\to0$. We will argue that this divergence is cancelled by a divergence of the one-loop fluctuations of the string worldsheet. This is the first indication that treating the terms $S_\text{FT}$ and $\Gamma_\mathbb{K}$ separately leads to inconsistencies which are hard to resolve. As we will emphasize, these terms should be treated together to obtain a finite one-loop correction of the holographic Wilson loop.


\subsubsection{Second order fluctuations}

We now turn to the quantum fluctuations of the worldsheet fields around the classical configuration \eqref{classicalSol}. This leads to a Gaussian model whose partition function is formally expressed in terms of determinants. The evaluation of the determinants will be the subject of the next section, but here we summarize the structure of the second order action, the derivation is carried out in appendix \ref{app:lagrangian}.

The Gaussian bosonic fields consist of eight scalar modes that determine the fluctuations of the embedding of the worldsheet in the ten-dimensional geometry. Let $X_\text{cl}$ denote the classical embedding of the string in the ten-dimensional geometry defined in \eqref{classicalSol}. The scalar fluctuations around this solution can then be written as
\be
X^\mu = X_\text{cl}^\mu + \delta X^\mu = X_\text{cl}^\mu + E^\mu_{\hat \mu} \zeta^{\hat \mu}\,,
\ee
where $E^\mu_{\hat \mu}$ are the ten-dimensional vielbeine, $E^\mu_{\hat\mu} E^\nu_{\hat \nu}\delta^{\hat \mu\hat \nu} = G^{\mu\nu}$.
A priori there are ten scalar fields $\zeta^{\hat\mu}$ as well as the dynamical worldsheet metric, but in static gauge these are reduced to eight. 
For our diagonal ten-dimensional metric, this simply means that $\zeta^{\hat\sigma}$ and $\zeta^{\hat\tau}$ are set to zero. 
To underline the fact that we are only working with the transverse fluctuations we use $\zeta^a$ where now $a=1, \dots, 8$ (see appendix \ref{app:lagrangian}).

For a proper treatment of the theory, in place of static gauge one should use conformal gauge which keeps the conformal factor of the metric as well as the longitudinal modes active. 
In addition, one then has to properly treat the reparametrization $bc$-ghost system as explained in \cite{Drukker:2000ep,Forini:2015mca}. The proper treatment ultimately leads to the same result as the static gauge approach where these extra modes are effectively cancelled against the ghosts%
\footnote{Some care must be exercised when treating the ghost determinant and its cancellation against the longitudinal modes. The $bc$ ghost determinant must exclude the zero modes which correspond to conformal Killing vector which are already accounted for in the measure.}.
For this reason, in this paper we choose the simpler static gauge and refer the reader to \cite{Drukker:2000ep,Forini:2015mca} for a detailed account on the difference between the two gauges. 

The fermionic fields are eight just like the scalars. These originate as 32 real target space fermions. The $\kappa$-symmetry gauge fixing reduces the physical modes to 16, which then are mapped to 8 two-dimensional fermions which we denote by $\theta_a$. The combined second order action can be written as
\be
\label{sk}
S_{\mathbb{K}}  = \f{1}{4\pi \ell_s^2}\int\sqrt{\gamma} \left( \zeta^a{\cal K}_{ab}\zeta^b +  \bar\theta^a{\cal D}_{ab}\theta^b\right)\dd^2\sigma\,.
\ee
Here the bosonic operators are diagonal with degeneracies $(4,2,2)$:
\be
{\cal K}_{ab} = \text{diag}({\cal K}_{x},{\cal K}_{x},{\cal K}_{x},{\cal K}_{x} ,{\cal K}_{y},{\cal K}_{y},{\cal K}_{z},{\cal K}_{z})\,.
\ee
Explicitly the operators take the form
\be\label{op-K-Ktilde}
{\cal K}_a = \e^{-2\rho}\tilde {\cal K}_a\,,\qquad \tilde {\cal K}_a=-\partial_\sigma^2 - \partial_\tau^2 + E_a\,,
\ee
with
\be\label{theEs}
\begin{split}
E_x &= \partial_\sigma^2 \rho+(\partial_\sigma \rho)^2-1 = \f{7+8\cosh 2\sigma}{\sinh^2 2\sigma}\,,\\
E_y &=\f12 \partial_\sigma^2 \rho= \f{1+2\cosh 2\sigma}{\sinh^2 2\sigma}\,,\\
E_z &=\f{1-3h^2}{2}\partial_\sigma^2 \rho+h^2(\partial_\sigma\rho)^2-h^2= \f{1+2h^2+2(1-h^2)\cosh 2\sigma}{\sinh^2 2\sigma}\,.
\end{split}
\ee
The fermionic operators are likewise diagonal with all identical entries  ${\cal D}_{ab} = {\cal D} \delta_{ab}$ where
\be\label{op-D-Dtilde}
\mathcal{D} = \e^{-3\rho/2}\tilde{\mathcal{D}}\e^{\rho/2}\,,\qquad \tilde{\mathcal{D}}=i\slashed{\partial} + \tau_3 a + v
\ee
and
\be\label{a-v-ferm}
a = \f{hi}{2\cosh\sigma}\,,\quad v = \f{3 i}{2\sinh\sigma} \,.
\ee

The path integral measure is implicitly defined with respect to the norms
\be
\lVert\zeta\rVert^2 = \int\dd^2\sigma\sqrt{\gamma}\zeta^a\zeta^b\delta_{ab}\,, \qquad \lVert\theta\rVert^2 = \int\dd^2\sigma\sqrt{\gamma}\,\bar\theta^a \theta^b\delta_{ab}\,,
\ee
through
\be
1 = \int \big[ D\zeta\big]\e^{-\f1{4\pi\ell_s^2}\lVert\zeta\rVert^2}\,,\qquad 1 = \int \big[ D\theta D\bar\theta\big]\e^{-\f1{4\pi\ell_s^2}\lVert\theta\rVert^2}\,.
\ee
Using this we perform the Gaussian path integral
\be\label{Gamma-K}
\Gamma_{\mathbb{K}}= -\log\int \big[D\zeta D\theta D\bar\theta\big]\e^{-S_\mathbb{K}} = \f12 \log \f{(\det{\cal K}_x)^4(\det{\cal K}_y)^2(\det{\cal K}_z)^2}{(\det{\cal D})^8}\,.
\ee
The subject of next section is evaluating these determinants.

\section{One-loop partition function} 
\label{sec:oneloop}

The focus of this section is to evaluate the one-loop functional determinants \eqref{Gamma-K}. 
To this end, instead of evaluating the determinants of ${\cal K}_a$ and ${\cal D}$, we would like to compute the considerably simpler determinants of the tilded operators $\tilde{\cal K}_a$ and $\tilde{\cal D}$, cf. \eqref{op-K-Ktilde} and \eqref{op-D-Dtilde}. 
The operators  $\tilde{\cal K}_a$ and $\tilde{\cal D}$ are flat operators, since they are obtained from the operators  ${\cal K}_a$ and ${\cal D}$ by stripping off a conformal factor. 
This is equivalent to perform a Weyl rescaling, which is allowed only in a Weyl invariant theory. 
The computation of the Weyl anomaly and its relation to UV divergences of the one-loop partition function are discussed in the subsection \ref{anomaly}. 
In subsection \ref{phaseshiftmethod} we illustrate the main points in the computation of the one-loop determinants \eqref{Gamma-K} by means of the phase shift method.


\subsection{Weyl anomaly}
\label{anomaly}

As mentioned above, the Weyl invariance of the theory should allow us to perform Weyl rescalings of the metric, and thus to employ the flat operators $\tilde{\cal K}_a$ and $\tilde{\cal D}$ in the computation of the functional determinants \eqref{Gamma-K}. 
In particular, the absence of a Weyl anomaly is required for this procedure. 
Fortunately, consistency of string theory requires the total central charge $c$ to vanish and by it the Weyl anomaly. In this subsection we will shortly review the Weyl anomaly in string theory and argue that indeed for our particular setup the total anomaly vanishes. Essentially this a simple consequence of our background geometry being a consistent background of string theory. 
We will also review how it relates to the logarithmic divergences of the partition function.

The quantization of the string in general backgrounds, leads to a non-trivial trace of the energy momentum tensor. This can be parametrized as\footnote{This expression is relevant for the bosonic string and for the superstring when the fermions vanish on-shell.}
\be\label{stringyemtensor}
2 \ell_s^2 \langle {T_i}^i\rangle = \ell_s^2 \beta^\Phi R_\gamma + \beta_{\mu\nu}^G\partial_i X^\mu\partial^i X^\nu + \beta_{\mu\nu}^B\partial_i X^\mu\partial_j X^\nu\epsilon^{ij}\,.
\ee
Here we have introduced the Weyl anomaly functions $\beta$ which are computed in the $\alpha'=\ell_s^2$ expansion of string theory. The consistency of the theory relies on the fact that all $\beta$-functions vanish, eliminating the Weyl anomaly completely. It is proven to all orders in the $\alpha'$-expansion that, if $\beta_{\mu\nu}^G=\beta_{\mu\nu}^B=0$, then $\beta^\Phi$ is constant \cite{Callan:1989nz} and we get back the familiar expression\footnote{We use the convention $T_{ij} = -\f{4\pi}{\sqrt{\gamma}}\f{\delta S}{\delta \gamma^{ij}}.$}
\be
\langle {T_i}^i\rangle = -\f{c}{12}R_\gamma\,,\qquad \beta^\Phi = -\f{c}6\,,
\ee
where $c$ is the total central charge. In the RNS formulation of the superstring in flat space, the contribution of the worldsheet scalars and fermions adds up to $3 D/2$. This should be combined with the reparametrization ghosts with $c=-26$ and the superconformal ghosts with $c=11$ giving\cite{Polyakov:1981rd,Polyakov:1981re}
\be
c = \f32 (D-10)\,.
\ee
We conclude that the total Weyl anomaly vanishes in the critical dimension $D=10$. 

The cancellation of the Weyl anomaly in the GS string works slightly differently as reviewed in \cite{Drukker:2000ep} (see references therein for further details). The lack of worldsheet supersymmetry means there are no superconformal ghosts and we only have eight worldsheet fermions instead of 10. However, it is important to note that the GS fermions are really 2D scalars with the wrong statistics, their contribution to the conformal anomaly is subtle to compute, but effectively they contribute four times the naive expectation for a normal 2D fermion, due to the fact that they couple to the worldsheet metric as scalars would.%
\footnote{Unfortunately this is obscured in our expressions since we use static gauge.} 
Combining all contributions yields 
\be
c = D-10\,,
\ee
which, again, vanishes in the critical dimension.

The Weyl anomaly is closely related to logarithmic divergences in the partition function. This can be observed directly from the definition of the quantum energy-momentum tensor as a variation of the effective action with respect to the conformal factor. We use $\gamma_{ij} = \e^{2\rho} \delta_{ij}$, then
\be\label{QEMtensor}
\langle {T_i}^i\rangle_{\mathbb{K}} = \f{2\pi}{\sqrt{\gamma}}\f{\delta \Gamma_{\mathbb{K}}}{\delta \rho}\,,
\ee
and the right-hand-side can be expressed in terms of the DeWitt-Seeley coefficients that control logarithmic divergences \cite{Drukker:2000ep}. In particular
\be
\delta \log \det {\cal K} = -2 a_2(\delta \rho|{\cal K})\,,\qquad \delta \log \det {\cal D}^2 = -2 a_2(\delta \rho| {\cal D}^2)\,,
\ee
where $a_2(f|{\cal O})$ is the second DeWitt-Seeley coefficients for the operator ${\cal O}$ evaluated on a test function $f$.
Using this we can rewrite \eqref{QEMtensor} as
\be\label{EMtensSeeley}
\langle {T_i}^i\rangle_{\mathbb{K}} =  \f14\trace b_2({\cal D}^2)-\f12\trace b_2({\cal K})\,,
\ee
where $a_2$ and $b_2$ are related through
\be
a_2(f|{\cal O}) = \f{1}{4\pi} \int\sqrt{\gamma} f b_2({\cal O}) + \text{boundary terms}\,.
\ee
Notice that in \eqref{EMtensSeeley} we have implicitly assumed that the variation $\delta \rho$ vanishes on the boundary. The ``local'' DeWitt-Seeley coefficients in our conventions (which are the same as those of \cite{Cagnazzo:2017sny}) take the form
\be
b_2({\cal D}^2) = -\f1{6} R_\gamma + 2\e^{-2\rho}(v^2-a^2)\,,\qquad b_2({\cal K}) = \f1{6} R_\gamma -  \e^{-2\rho}E\,.
\ee
The complete expression for the quantum energy momentum tensor for the eight scalars and eight fermions is then given by 
\be\label{traceanomalydef}
\langle {T_i}^i\rangle_{\mathbb{K}}=-\f12\Big(2R_\gamma -\e^{-2\rho}\trace E - \e^{-2\rho}\trace(v^2-a^2)\Big)\,,
\ee
where the trace should be understood as over all bosonic and fermionic masses \eqref{theEs}-\eqref{a-v-ferm}. 

Just as in \eqref{stringyemtensor}, the terms in \eqref{traceanomalydef} should be separated into terms proportional to $R_\gamma$ on one hand, and $\partial_i X^\mu \partial_j X^\nu$ on the other. 
However, since we have worked in static gauge, it is difficult to separate the two terms. 
In order to do so, we would have to expand the Polyakov action using the conformal gauge and a background metric that is not identified with the induced metric (see \cite{Drukker:2000ep,Forini:2015mca} for a detailed discussion). 
In addition to the eight transverse scalars, we would now have the two longitudinal modes as well as ghost fields. The contribution to $\langle {T_i}^i\rangle$ that is proportional to $\partial_i X^\mu \partial_j X^\nu$ turns out to be the total mass contribution of all physical fields. The total scalar masses of all ten fields is $\e^{-2\rho}\trace E-R_\gamma$ while the fermions still give $\e^{-2\rho}\trace (v^2-a^2)$. Note that including also the ghost fields, the total contribution to the DeWitt-Seeley coefficients is the same in the two gauges~\cite{Drukker:2000ep}. 
With this in mind we can suggestively rewrite \eqref{traceanomalydef} as
\be\label{Tseparated}
\langle {T_i}^i\rangle_{\mathbb{K}}=-\f12\Big(R_\gamma -\e^{-2\rho}\trace E - \e^{-2\rho}\trace(v^2-a^2)\Big) -\f12 R_\gamma\,,
\ee
where the first term should now be understood as the one proportional to $\partial_i X^\mu \partial_j X^\nu$.

Indeed we can explicitly verify using the ten-dimensional solution that
\be
R_\gamma -\e^{-2\rho}\trace E=  \partial_i X^\mu \partial^i X^\nu\Big[ R_{\mu\nu} -\f12 |H|^2_{\mu\nu} \Big]\,,
\ee
which are the first two terms in the expected Weyl anomaly function $\beta^G_{\mu\nu}$. Similarly we have checked that
\be
-\e^{-2\rho}\trace(a^2-v^2)=-\f14 \partial_i X^\mu \partial^i X^\nu\e^{2\Phi}\sum_n|F_n|^2_{\mu\nu} \,.
\ee
Here the sum over $n$ runs over form fields as well as their Hodge duals (this is the so-called democratic formulation). Using these results we see that the $\partial_i X^\mu \partial_j X^\nu$-terms in \eqref{traceanomalydef} can be expressed as
\be\label{missingterm}
-\f12\partial_i X^\mu \partial^i X^\nu\Big[ R_{\mu\nu} -\f12 |H|^2_{\mu\nu} -\f14 \e^{2\Phi}\sum_n|F_n|^2_{\mu\nu}\Big]=\partial_i X^\mu \partial^i X^\nu\nabla_\mu\nabla_\nu \Phi\,,
\ee
where we used the equations of motion of ten-dimensional supergravity. We therefore see that if the dilaton is non-constant (as for our background) then the $\partial_i X^\mu \partial_j X^\nu$-terms in the energy momentum tensor do not vanish as they must for a consistent theory.

This should not come as a particular surprise since we have neglected to take into account the classical Weyl ``anomaly'' of the  FT action. It is well known that the classical Weyl rescaling of the FT action is  cancelled by the anomaly of the one-loop fluctuations of the string (see for example \cite{Callan:1989nz}). We can correct for this by adding to \eqref{traceanomalydef} the classical energy-momentum tensor computed using the FT action \eqref{FTAction}
\be
({T_i}^i)_\text{FT} = -\partial_i X^\mu \partial^i X^\nu\nabla_\mu\nabla_\nu \Phi\,,
\ee
which exactly compensates for the missing term in \eqref{missingterm}, ensuring that the $\partial_i X^\mu \partial_j X^\nu$-terms in the full energy momentum tensor do in fact vanish as expected.

We are then left with 
\be\label{finalEMtens}
\langle {T_i}^i\rangle =\langle {T_i}^i\rangle_{\mathbb{K}} + ({T_i}^i)_\text{FT} = -\f12 R_\gamma\,,
\ee
which is identical to the result one would get from an similar treatment of the GS string in flat space \cite{Drukker:2000ep}. The remaining anomaly is universal and is cancelled here in exactly the same way as in flat space. Roughly speaking this is accomplished by the combination of two effects. First, the transformation of the GS fermions to two-dimensional fermions on the worldsheet is accompanied by a Jacobian which contributes additional $-R_\gamma$, next, in conformal gauge the FP determinant can be rewritten as a $bc$-ghost system for which zero-modes must be excluded. This produces an additional $(3/2) R_\gamma$ which makes the total conformal anomaly vanish (see for example \cite{Blumenhagen:2013fgp}).

Since we will not take care of these two ingredients, and only use its universal nature \cite{Giombi:2020mhz}, our partition function will carry logarithmic divergences controlled by the DeWitt-Seeley coefficients. In terms of the Weyl anomaly, the divergence is just
\be
\f1{2\pi} \int \langle {T_i}^i\rangle \vol_\gamma = -\chi = -1\,,
\ee
where $\chi$ is the Euler characteristic of the worldsheet. 
We will recover exactly this logarithmic divergence when we explicitly compute the partition function in section \ref{phaseshiftmethod}.

To summarize this subsection, our treatment of the Weyl anomaly shows that because the ten-dimensional dilaton is non-trivial in our background, we should not separately compute $S_\text{FT}$ and $\Gamma_\mathbb{K}$ if we want to perform our desired Weyl rescaling. Rather these should be treated as a combined object
\be
W \equiv S_\text{FT}+\Gamma_\mathbb{K}\,.
\ee
Since the \emph{total} Weyl anomaly vanishes in string theory, we can perform the Weyl transformation to the flat metric as desired. Due to the fact that we do not carefully keep account of all string theory ingredients discussed above, our one-loop determinants will carry divergences. These divergences are however argued to be universal. As we will explain in further detail in section \ref{ratio}, in order to control the universal contributions we suggest to compute the ratio of two string partition functions keeping in mind that all universal factors drop out.

For now, let us discuss a different issue that we encounter. We denote the corresponding flat space quantities (i.e. Weyl transformed quantities) by a tilde. It turns out that $\tilde S_\text{FT}$ identically vanishes. This is somewhat surprising since we expect to find something in the spirit of $\chi \log g_s$. 
Since our worldsheet is a disc (that is $\chi=1$), and since the vacuum value of the dilaton does not vanish, the FT term should not vanish.

The problem is that the Weyl transformation effectively changes the topology of our manifold due to the choice of coordinates used. The ``center'' of our worldsheet is located in our coordinates at $\sigma\to\infty$. In the Weyl rescaled flat metric, this point is pushed infinitely far away. 
This was first emphasized in \cite{Cagnazzo:2017sny}. 
In all practical computations we must place an IR cutoff at some large finite $\sigma=R$ which changes the topology of the worldsheet to a cylinder. The Euler characteristic of the cylinder vanishes, which explains why the direct application of the formula \eqref{FTAction} results in the answer $\tilde S_\text{FT}=0$. What we have neglected to take into account is the contribution of the small disc located at $\sigma\ge R$ that we cut off by the Weyl rescaling. In this region the dilaton \eqref{dilaton} is approximately constant,  and we can just use 
\be
\label{S-FT-final}
\tilde S_\text{FT} =\chi\lim_{\sigma\to\infty} \Phi_0 = -\log \f{N\pi}{\xi^{3/2}}\,.
\ee
In principle we should also retain a contribution from one-loop fluctuations of fields inside this small disc. However, the bosonic and fermionic operators are exactly free in this limit and so we obtain only the universal contribution due to UV divergences. We will take these into account when computing the one-loop fluctuation for $\sigma \le R$ and so do not account for these here. In total we then have
\be
W = -\log \f{N\pi}{\xi^{3/2}} + \tilde\Gamma_{\mathbb K}(R)\,,
\ee
where $\tilde\Gamma_{\mathbb K}(R)$ is the one-loop partition function of the Weyl rescaled operators using an IR cutoff at $\sigma = R \gg 1$. Notice that this expression, and in particular the FT term \eqref{S-FT-final} is free from the divergence encountered in \eqref{FTtermdivergent}. We expect that if we had not performed the Weyl rescaling discussed here, the one-loop fluctuations $\Gamma_{\mathbb K}(R)$ would carry a similar divergence as \eqref{FTtermdivergent} and cancel it. However, using the Weyl rescaled quantities both quantities are now free from this powerlaw UV divergence.

\subsection{Phase shift method}
\label{phaseshiftmethod}

Our remaining task is to compute the one-loop partition function $\tilde\Gamma_{\mathbb K}(R)$  for the tilded (flat) operators, that is
\be
\label{Gamma-tilde}
\tilde\Gamma_{\mathbb K}(R)= \f12 \log \f{(\det{\tilde{\cal K}}_x)^4(\det{\tilde{\cal K}}_y)^2(\det{\tilde{\cal K}}_z)^2}{(\det{\tilde{\cal D}})^8}\,,
\ee
where the operators are given in equations \eqref{op-K-Ktilde} and \eqref{op-D-Dtilde}. 
To this end we will use the phase shift method \cite{Chen-Lin:2017pay,Cagnazzo:2017sny}.
The operators we are interested in  are two-dimensional, for example the bosonic operators are of the form
\be
\tilde {\cal K}_a=-\partial_\sigma^2 - \partial_\tau^2 + E_a (\sigma)\,,
\ee
where the potentials $E_a$ are defined in \eqref{theEs}. 
The first step is to Fourier expand with respect to the angle $\tau$, then $\partial_\tau$ is mapped to $i \omega$. 
Bosonic and fermionic operators obey periodic and anti-periodic boundary conditions with respect  to the $\tau$ coordinate, respectively, then $\omega$ will be an integer for the bosonic fluctuations and half-integer for the fermionic ones. 
Computing the functional determinants in \eqref{Gamma-tilde} amounts to solve the spectral problem for these now one-dimensional Schr\"odinger operators, i.e.
\be\label{eigenvaluedef}
\tilde {\cal K}_a \,\eta_{ \omega}(\sigma) =\left(-\partial_\sigma^2 +\omega^2+ E_a (\sigma)\right) \eta_{ \omega}(\sigma)= \lambda\, \eta_{\omega}(\s)\,,
\ee
where $\eta_{\omega}(\s)$ represents now only the ``radial'' component of the full wave function, that is $\Psi(\s, \tau)=\Sigma_\omega e^{i \omega \tau} \eta_{\omega}(\s)$. 
For large $\sigma$ the potentials $E_a(\sigma)$, $v(\sigma)$ and $a(\s)$ asymptote to zero, and we are left with free operators. 
Hence, the solutions asymptotically behave as waves. 
The effect of the potential (and hence the information about the spectrum) is contained in a phase shift, $\delta$, which measures at large $\s$ how close the solution is to a free (ingoing or outgoing) wave, that is 
\be
\label{eta-asymptotic}
\eta_{\omega} \to C \sin(p\sigma + \delta(\omega,p))\,.
\ee
From here it is also manifest that the dispersion relation is simply
\be\label{disp-rel}
\lambda= \omega^2+p^2\,. 
\ee
The goal is to compute the phase shift $ \delta(\omega,p)$ for each of our bosonic and fermionic operators.
This scattering problem is clearly illustrated in~\cite{Cagnazzo:2017sny}, hence here we only summarize the main steps.

Firstly, our fluctuations obey Dirichlet boundary conditions at $\s=0$. 
In the UV limit, the potentials diverge $E_a \sim \sigma^{-2}$ which implies that the wave functions are either non-normalizable or they vanish. We will choose normalizable wavefunctions that vanish in the UV:
\be\label{bc-Dirichlet-UV}
\eta_{\omega}(\sigma =0)=0\,.
\ee

Secondly, to get a discrete spectrum we introduce an IR cutoff at large distance $\sigma = R$ and impose Dirichlet boundary conditions at $\s=R$, that is $\eta_{\omega}(R)=0$. 
Given the asymptotic behaviour \eqref{eta-asymptotic} of the solutions,  this quantization condition reads
\be\label{quantization}
pR + \delta(\omega,p) = \pi k\,,
\ee
where $k$ is a positive integer.
From this we can read the density of states, or the multiplicity of the eigenvalues,
\be
\rho = \f{\dd k}{\dd p} = \f{1}{\pi}\left(R + \f{\dd \delta(\omega,p)}{\dd p}\right)\,.
\ee

Hence, the functional determinant reduces to
\be
\log \det \tilde {\cal K} = \sum_\omega \int_0^\infty \f{\dd p}{\pi}\left(R + \f{\dd \delta(\omega,p)}{\dd p}\right)\log (p^2+\omega^2)\,,
\ee
where we used the fact that the spectrum is approximately continuous in the large $R$ limit, and the explicit form of the eigenvalues \eqref{disp-rel}.  
We can then integrate by parts over $p$, and replace the sum over Matsubara frequencies with a contour integral over $\omega$, which gives contributions only at the poles $\omega=\pm i p$ and hence $\lambda=0$.%
\footnote{More details can be found in~\cite{Cagnazzo:2017sny}.} 
This gives for the bosonic operators
\be
\log \det \tilde {\cal K} =- \int_0^\infty \dd p\,\coth (\pi  p)\Big( 2pR + \delta(i p,p)+ \delta(-i p,p) \Big)\,,
\ee 
and for the fermionic operators
\be
\log \det \tilde {\cal D} =- \int_0^\infty \dd p\,\tanh (\pi  p)\Big( 2pR + \delta(i p,p)+ \delta(-i p,p) \Big)\,.
\ee 

Finally, we can use the fact that the bosonic operators \eqref{op-K-Ktilde} are Hermitian, which leads to $\delta(i p,p)= \delta(-i p,p)$. We will denote the phase shifts corresponding to the bosonic operators by simply $\delta_a$ where $a=x,y,z$. 
On the other hand, the fermionic operators \eqref{op-D-Dtilde} are not Hermitian,  and so we will have two independent phase shifts for $\omega=\pm i p$, which we  denoted by $\delta_\pm$. 

At this point we are ready to write the full expression for the effective action $\tilde\Gamma_{\mathbb K}(R)$ \eqref{Gamma-tilde} in terms of the phase shifts, that is collecting all the terms we have 
\be\label{phaseshiftformula}
\tilde\Gamma_{\mathbb K}(R)= -\int_0^\infty \dd p\, \Big[\coth (\pi  p)(4\delta_x+2\delta_y+2\delta_z)-\tanh (\pi  p)(4\delta_{+}+ 4\delta_{-})\Big] - R\,,
\ee
where we have performed the explicit $p$-integral multiplying the cutoff $R$.

\subsubsection{Phase shifts for the bosonic operators}

In this subsection we focus on the bosonic operators. 
As we have seen above, in order to compute the phase shifts, we have to solve the zero eigenvalue Schr{\"o}dinger problem
\be\label{bosoniceq}
\tilde{\cal K}_a \eta_a(\s) = 0\,, \qquad a=x, y,z\,, 
\ee 
with boundary conditions \eqref{bc-Dirichlet-UV} and \eqref{quantization}. 
Since the bosonic operators are Hermitian, the two independent solutions come in complex conjugate pairs $\eta_{a}(p;\sigma)$ and $\bar \eta_a(p;\sigma)$. Like \cite{Cagnazzo:2017sny}, we normalize our basis functions such that
\be\label{normalization-eta}
\lim_{\sigma\to\infty}\e^{-ip\sigma}\eta_a(p;\sigma) = 1 = \lim_{\sigma\to\infty}\e^{ip\sigma}\bar \eta_a(p;\sigma)\,.
\ee
The explicit form of the bosonic basis function is then
\be
\label{eta-bos-h1}
\begin{split}
\eta_{x}(p;\sigma) &= \big(2\sinh \sigma\big)^{i p}(\coth\sigma)^{1/2}\, \HyperGeo\Big(-\tfrac{1+i p}{2},\tfrac{3-ip}{2};1-ip; -{\rm csch}^2\s \Big)\,, \\ 
\eta_{y}(p;\sigma) &= \big(2\sinh \sigma\big)^{i p}(\coth\sigma)^{1/2}\, \HyperGeo\Big(-\tfrac{i p}{2},\tfrac{2-ip}{2};1-ip;-{\rm csch}^2\s \Big)\,, \\ 
\eta_{z}(p,h=1;\sigma) &= \big(2\sinh \sigma\big)^{i p}(\coth\sigma)^{-1/2}\, \HyperGeo\Big(\tfrac{1-i p}{2},-\tfrac{1+ip}{2};1-ip;-{\rm csch}^2\s \Big)\,, \\ 
\end{split}
\ee
for the three operators $\tilde{\cal K}_x$, $\tilde{\cal K}_y$, and $\tilde{\cal K}_z$. 
Here we have set the parameter $h$ equal to 1, however the results for generic values of $h$ are discussed in appendix \ref{app:res-various-h}.  
A wave function that is regular at $\sigma\to0$ vanishes there, and can be directly constructed as follows
\be
\eta(p;\sigma) = {\cal N}\big(\bar\eta(p;0)\eta(p;\sigma)-\eta(p;0)\bar\eta(p;\sigma)\big)\,,
\ee
where ${\cal N}$ is an unimportant normalization of the wave function. The phase shift can then be determined directly by evaluating the limit $\sigma\to\infty$ and imposing the quantization condition \eqref{quantization}. The resulting phase shifts for the three bosonic operators for $h=1$ are
\be
\label{delta-bos-h1}
\begin{split}
\delta_x(p) &= \text{Arg}\, \Big[{2^{-ip}\Gamma(\tfrac{3-ip}{2})^2\Gamma(1+ip)}\Big]\,,\\
\delta_y(p) &= \text{Arg}\, \Big[\Gamma(1-\tfrac{ip}{2})\Gamma(\tfrac{1+ip}{2})\Big]\,,\\
\delta_z(p) &= \text{Arg}\, \Big[{2^{-ip}\Gamma(\tfrac{1-ip}{2})\Gamma(\tfrac{3-ip}{2})\Gamma(1+ip)}\Big]\,.
\end{split}
\ee
The bosonic phase shifts for any $h$ are reported in appendix \ref{app:res-various-h}.

\subsubsection{Phase shifts for the fermionic operators}
\label{sec:phase-shift-ferm}

Here, we start by illustrating the computation and the results of the fermionic phase shifts. At the end of the section we collect our findings in the expression \eqref{final-gamma-tilde}.

Unfortunately, for the fermionic operators we could not analytically solve the wave equation for all $h$. 
We are now looking at the following matrix equation
\be\label{fermioniceq}
\tilde{\mathcal{D}}\eta =0\,,
\ee
where the operator $\tilde{\mathcal D}$ is given in \eqref{op-D-Dtilde}. 
Only for a few values of $h$ ($h=0, 3$) we can find an analytic solution to the above equations, and the results are reported in appendix \ref{app:res-various-h-ferm}.  
For this reason, we have to resort to numerics. 
Exactly as for the bosonic operators, we impose regular boundary conditions \eqref{bc-Dirichlet-UV} and read off the oscillating wave function for large $\sigma$, cf. \eqref{eta-asymptotic}. We must do this for both $\omega = i p$ and $\omega=- i p$ since the fermionic operator \eqref{op-D-Dtilde} is not Hermitian, and thus, the corresponding phase shifts will not be equal. 

\begin{figure}[h]
\centering
\includegraphics[width=0.8\textwidth]{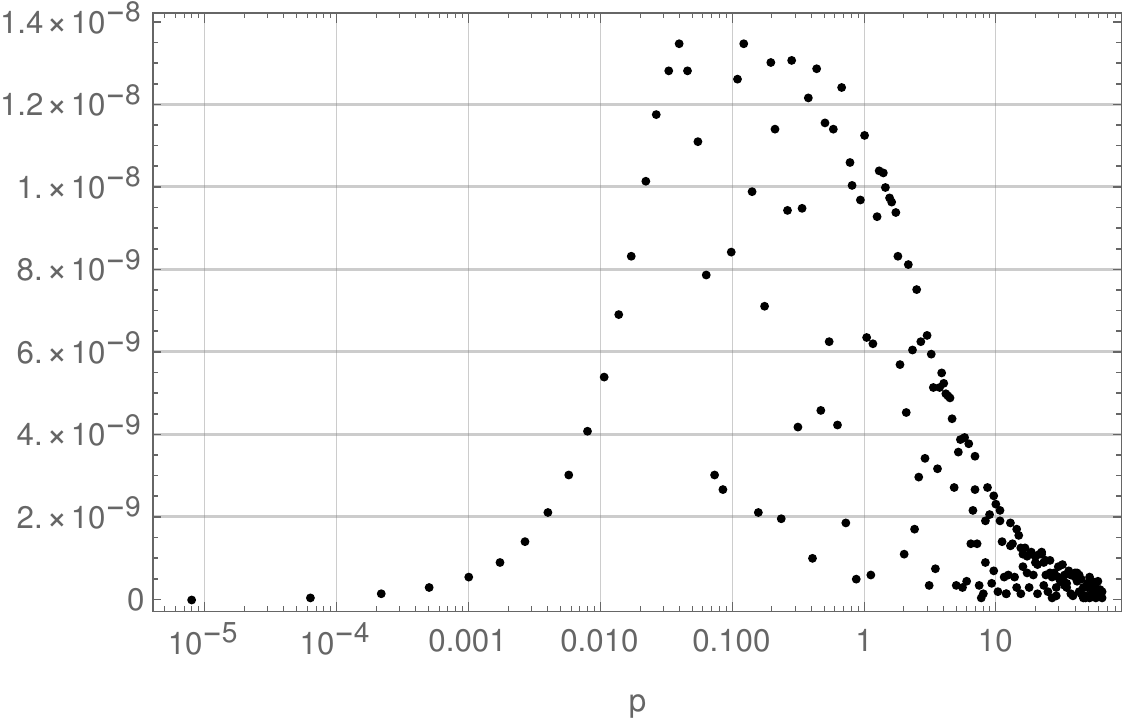}
\caption{\label{errors}The total absolute numerical error for the sum of all fermionic phase shifts computed using numerical methods compared with the analytic expressions obtained for $h=0, h=3$, and reported in \eqref{delta-ferm-h0} and \eqref{delta-ferm-h3}.}
\end{figure}

The most stable approach we have found to numerically  extract the phase shifts is to first compute the two component wave function to high accuracy, imposing regular boundary conditions at small $\sigma$.
Then, we evaluate a particular ratio of these two components at large $\s$ which approaches a constant, which is just the phase shift. 
At large $\sigma$ the wave functions take the form\footnote{We reuse $\eta$ here for a two-dimensional fermionic wave function.}
\be
\eta(\sigma) =\begin{pmatrix}\eta_1(\s) \\\eta_2(\s) \end{pmatrix}\sim \begin{pmatrix}c_1 \e^{\mp i p \sigma}\\ c_2 \e^{\pm i p\sigma}\end{pmatrix}\,,
\ee
where upper sign refers to $\omega=+i p$, and lower sign to $\omega=-i p$. Here $c_{1,2}$ are two constants that depend on $p$ and must be computed numerically. Using this asymptotic form we can impose quantization condition at large $\sigma=R$ which takes the form \cite{Cagnazzo:2017sny} (see also \cite{Medina-Rincon:2019bcc} for a more detailed discussion)
\be
\tau_2 \eta(R) = \eta(R)\,\quad \text{or} \quad  p R = k\pi  \pm\f{1}{2} \text{Arg} \f{i c_1}{c_2}\,.
\ee
Comparing with \eqref{quantization} and using the components of the wave function we find
\be
\delta = \mp\f{1}{2} \text{Arg} \Big(\f{i \e^{\pm 2i pR}\eta_1(R)}{\eta_2(R)}\Big)\,,
\ee
where the signs are correlated with $\omega=\pm i p$.
We have performed many numerical evaluations for a large range of $p$ and for many values of the parameter $h$, as discussed in appendix \ref{app:phaseshift}. 
For $h=0$ and $3$ we can  compare directly the numerical results with the analytic answers \eqref{delta-ferm-h0}-\eqref{delta-ferm-h3} to evaluate the precision of our code. We find excellent agreement with the analytic phase shifts with errors ranging between $10^{-9}$ and $10^{-7}$ (see figure \ref{errors}). 
We also compare our numerical results against a WKB approximation at large $p$ finding a perfect match.  

\begin{figure}[h]
\centering
\includegraphics[width=0.8\textwidth]{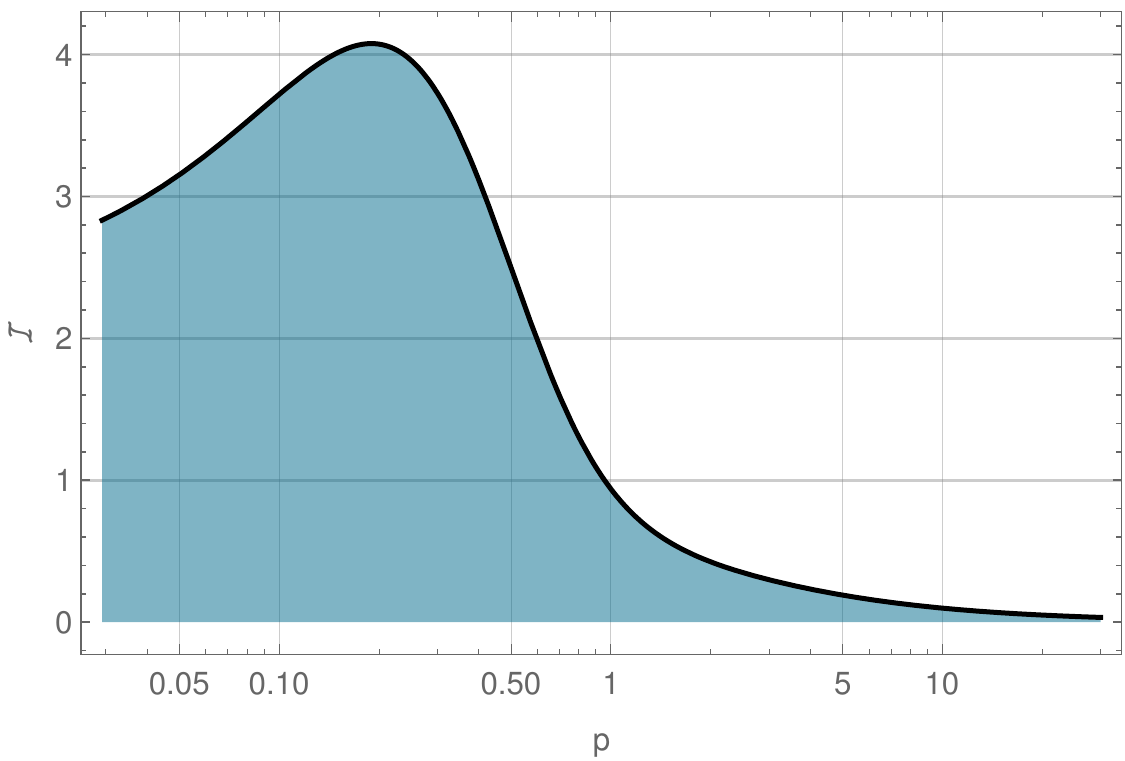}
\caption{\label{integrand}The full integrand in \eqref{phaseshiftformula} obtained by combining numerical results for the fermionic phase shifts and the analytic expression for the bosonic phase shift. We have set the parameter $h=1$ in this figure. The integral over the shaded region gives  us the regularized $\Gamma_{\mathbb K}$.}
\end{figure}

\vskip 0.3 cm
The numerical fermionic phase shifts are combined with the analytic ones for the bosons \eqref{delta-bos-h1} into the integrand in \eqref{phaseshiftformula} denoted by $\mathcal{I}(p)$ and shown in figure \ref{integrand} for $h=1$. 
We then perform a numerical integration of our integrand ${\cal I}(p)$. 
This is logarithmically divergent for large $p$ with a coefficient given by the worldsheet Euler characteristic $\chi=1$, as we expected from our analysis in section \ref{anomaly}. 
For this reason, we numerically integrate up to a large UV cutoff $p=\Lambda$ and subtract $\log\Lambda$ from the numerical answer. 
At this point the integral is finite, and we find that it matches $2\log \pi$ up to five digits. 
We therefore conclude that, for $h=1$, the one-loop effective action \eqref{Gamma-tilde} is given by
\be\label{final-gamma-tilde}
\tilde\Gamma_{\mathbb K}(R) =  2\log \pi +\log(\Lambda \e^{-R})\,. 
\ee

\section{Ratio with ABJM and match with QFT}
\label{ratio}

In this section we collect the results obtained so far. 
Up to one-loop in the 't Hooft coupling constant $\xi$, the logarithm of the string partition function \eqref{log-Z-tocompute} comprises of three terms: the classical regularized action \eqref{s-classical}, the Fradkin-Tseytlin contribution \eqref{S-FT-final}, and finally the contribution coming from the fluctuations \eqref{final-gamma-tilde}. 
Hence, collecting all the pieces, we find that the string partition function takes the form
\be
\label{logZD4result}
\log Z_\text{SYM}^\text{string} \approx \xi +\log \f{N_\text{SYM}}{\xi^{3/2}\pi} - \log(\Lambda \e^{-R_\text{SYM}})\,,
\ee
where $R_{\rm SYM}$ is the IR cutoff on the coordinate $\sigma$.
We have introduced the label SYM on the integer $N$ and the cutoff $R$ in order not to confuse with ABJM quantities which we will encounter in this section.

In order to compare this result with the field theory prediction \eqref{W5-final} we must first find a way to deal with the two cutoffs $\Lambda, R_{\rm SYM}$. 
We suggest to follow a similar procedure used in \cite{Forini:2015bgo,Faraggi:2016ekd,Forini:2017whz,Cagnazzo:2017sny,Medina-Rincon:2018wjs}, that is to compute a ratio of string partition functions, where the string worlsheets have the same topology, and then compare this with the corresponding ratio of Wilson loop expectation values on the field theory side. 
In \cite{Forini:2015bgo,Faraggi:2016ekd,Forini:2017whz,Cagnazzo:2017sny,Medina-Rincon:2018wjs} the ratio considered was that of a latitude Wilson loop with the circular one.
Without having an analogue of the latitude WL in 5D SYM, this approach seems not applicable. 
However, as we have anticipated in the introduction and discussed in section \ref{sec:oneloop}, the divergences plagued by our string theory computation are universal. Indeed, the UV divergent piece is present even for a string in flat space \cite{Drukker:1999zq}, whereas the IR divergence is directly related to the procedure we used to compute the one-loop functional determinants.
Therefore, we should be able to compute a ratio of string partition functions (for  string worldsheets with the same topology), and cancel the two regulators as long as the same computational method is used, implying that the same regularization scheme is employed. 

Staying within a type IIA/M-theory setup, we will compute the ratio of our string partition function \eqref{logZD4result} with that of a circular string in AdS$_4\times {\bf C}P^3$. 
The partition function of this string should capture the expectation value of a $1/2$-BPS circular (fermionic) Wilson loop operator~\cite{Drukker:2009hy} in the ABJM theory
\cite{Aharony:2008ug} which can be computed by supersymmetric localization \cite{Kapustin:2009kz,Drukker:2008zx,Marino:2009jd,Drukker:2010nc}
\be
\label{WABJM}
\langle{\cal W} \rangle_\text{ABJM}  \approx \f{N_\text{ABJM}}{4\pi\lambda}\e^{\pi\sqrt{2\lambda}}\,,\qquad \lambda\gg 1\,.
\ee
Note that here we have used the conventions of \cite{Giombi:2020mhz} when normalizing the Wilson loop VEV, and included the rank of the gauge group $N_\text{ABJM}$ in this expression.

Let us now consider the dual string wrapping the equator of $\sphere^3$ inside AdS$_4$ (or its analytic continuation ${\bf H}^4$) in global coordinates. The classical solution was firstly discussed in \cite{Drukker:2008zx,Chen:2008bp}. 
Using the same conventions as we have done so far, the metric on the string worldsheet is in this case given by the conformal factor
\be\label{M2metric}
\e^{2\rho } = \f{\pi\sqrt{2\lambda}\,\ell_s^2}{\sinh^2\sigma}\,.
\ee
where $\lambda$ is  the corresponding 't Hooft coupling.%
\footnote{In the AdS$_4$/CFT$_3$ duality the 't Hooft coupling is given by $\lambda={N\over k}$ where $k$ is the level of gauge groups $\grU(N)_k\times \grU(N)_{-k}$~\cite{Aharony:2008ug}.} 
This allows us to compute the regularized classical action
\be
\label{ads4-cl}
S_\text{classical} = -\pi\sqrt{2\lambda}\,,
\ee
reproducing the exponential behavior of the Wilson loop vev in \eqref{WABJM}.

Next we compute the one-loop correction to the classical result. This is exactly the same procedure we have been doing so far for the D4-branes except that, since the geometry is AdS, it is technically simpler. 
The one-loop correction consists of two terms: The FT action and the one-loop fluctuations of worldsheet fields.
For the AdS$_4$ string the FT action is simple to evaluate due to the fact that the dilaton is constant
\be\label{ads4-SFT}
S_\text{FT} = \chi \log g_s = -\log \f{N}{\sqrt{\pi}(2\lambda)^{5/4}}\,,
\ee
where the string coupling is indeed $g_s =\left(32 \pi^2 \lambda^5\right)^{1/4} N^{-1}$. Here we have simply followed the steps discussed at the end of section \ref{anomaly}. 

The one-loop fluctuation of worldsheet fields is computed in much the same way as in section \ref{phaseshiftmethod}. 
First, we perform a Weyl transformation to strip off the metric factor so that we can compute the functional determinant of operators on flat space. 
Next, we use the phase shift method to compute determinants themselves (see appendix \ref{app:Ads}).

The one-loop partition function for a string in AdS$_4\times \mathbf{C} P^3$  dual to a 1/2-BPS circular (fermionic) WL was computed in~\cite{Kim:2012tu, Aguilera-Damia:2018bam}, where either a Gel’fand-Yaglom method or a heat kernel approach were utilized.%
\footnote{See also  \cite{Buchbinder:2014nia,Giombi:2020mhz} for a computation of the one-loop string partition function directly using the heat kernel method without performing any Weyl rescaling. }
These methods implicitly employ a different UV regularization scheme with respect to the phase shift method used here, and so we cannot simply borrow their results. In fact, for this reason, the answer \eqref{GammaAdS4} we obtain does not agree e.g. with the heat kernel result.%
 \footnote{See eq. (2.21) in~\cite{Giombi:2020mhz} for a summary of the regularised one-loop string effective action in AdS$_n$ calculated by means of the heat kernel.}
In \cite{Medina-Rincon:2019bcc} the phase shift method was used to calculate the 1/2-BPS circular WL in ABJM at one loop at strong coupling.  
Since there the focus was slightly different, and also in order to match with the notation of this work, we have included our calculation of the AdS$_4$ string partition function in appendix \ref{app:Ads}. 
The final expression is
\be
\label{GammaAdS4}
\Gamma_{\text{AdS}_4} = 2 \log \pi +\log(\Lambda \e^{-R_\text{ABJM}})\,,
\ee
where $R_{\rm ABJM}$ above is an IR cutoff in the coordinate $\sigma$. 
Combining \eqref{ads4-cl}, \eqref{ads4-SFT}, and \eqref{GammaAdS4}, the string partition function for the AdS$_4$ string is
\be
\label{logZ-ads4-notinv}
\log Z_\text{ABJM}^\text{string} \approx \pi\sqrt{2\lambda}  +\log \f{N}{(2\pi^2 \lambda)^{5/4}} -\log(\Lambda \e^{-R_\text{ABJM}})\,.
\ee
It is important to note that the IR cutoff appearing in \eqref{GammaAdS4}, cannot be identified with the IR regulator $R$ in \eqref{logZD4result}, as indeed underlined by the different suffix. 
Since the metric is different for the two cases, it is not sensible to identify the two cutoffs. 
Instead, we should follow the procedure in \cite{Cagnazzo:2017sny} and replace $R$ by a diffeomorphism invariant regulator given by the area of the worldsheet that is being cut off in the computation of the phase shifts. 
This dimensionless area is defined by
\be
A = \f{2\pi}{\ell_s^2}\int_R^\infty \e^{2\rho}\dd\sigma\,,
\ee
for the two cases at hand, i.e. \eqref{D4metric} and \eqref{M2metric}, we obtain
\be
\label{def-A}
A_\text{ABJM} = 4\pi^2\sqrt{2\lambda}\,\e^{-2R_\text{ABJM}}\,,\qquad A_\text{SYM} = 16\pi\xi\,\e^{-2R_\text{SYM}}\,.
\ee
The cutoffs defined in terms of the area $A$ can now safely be identified for the two cases, that is
\be
A_\text{ABJM} = A_\text{SYM} \equiv A\,.
\ee

We can now express the string partition function in the two cases,  i.e.\eqref{logZD4result} and \eqref{logZ-ads4-notinv}, using the diffeomorphism invariant cutoff $A$ defined in \eqref{def-A}, then we have 
\be
\begin{split}
\log Z_\text{SYM}^\text{string} &\approx \xi +\log \f{ 4 N_\text{SYM}}{\xi\sqrt{\pi}} - \log(\Lambda \sqrt{A})\,,\\
\log Z_\text{ABJM}^\text{string} &\approx \pi\sqrt{2\lambda} +  \log \f{ N_\text{ABJM}}{\pi^{3/2}\lambda} - \log (\Lambda \sqrt{A})\,.
\end{split}
\ee
Since the cutoffs in these two expressions are now the same we can safely cancel them in a ratio of string partition functions, and obtain 
\be
\f{Z_\text{SYM}^\text{string}}{Z_\text{ABJM}^\text{string}} = \Big(\f{N_\text{ABJM}}{4\pi\lambda}\e^{\pi\sqrt{2\lambda}}\Big)^{-1}\Big( \f{N_\text{SYM}}{\xi} \e^{\xi}\Big)\,.
\ee
Let us then consider the ratio of the two localization results for the vev of the 1/2-BPS Wilson loop operators, expanded at strong coupling,  in five-dimensional SYM \eqref{WL-strong-coupling-exp} on one hand,  and ABJM \eqref{WABJM} on the other. It is then clear that comparing the two ratios we find a perfect match:
\be
{\langle{\cal W} \rangle_\text{SYM}\over \langle{\cal W} \rangle_\text{ABJM}}= \f{Z_\text{SYM}^\text{string}}{Z_\text{ABJM}^\text{string}} \,. 
\ee

\bigskip
\bigskip
\bigskip
\leftline{\bf Acknowledgements}
\smallskip
\noindent We are grateful to Pieter Bomans, Nikolay Bobev, Nadav Drukker, Valentina Forini, Luca Griguolo, Daniel Medina-Rincon, Joe Minahan, Domenico Seminara, L{\'a}rus Thorlacius, Maxime Tr{\'e}panier, Arkady Tseytlin, Edoardo Vescovi, and Kostya Zarembo for useful discussions. We especially acknowledge Domenico Seminara, Arkady Tseytlin and Kostya Zarembo for comments on the manuscript. FFG is supported by the University of Iceland Recruitment Fund. VGMP is partially supported by grants from the University of Iceland Research Fund. FFG and VGMP are supported by the Icelandic Research Fund under grant 228952-051.

\newpage
\appendix
\section{One-loop lagrangian}
\label{app:lagrangian}
In this appendix we summarise the derivation of the bosonic and fermionic operators $\mathcal{K}$ \eqref{op-K-Ktilde} and $\mathcal{D}$ \eqref{op-D-Dtilde}, respectively.

\subsection{Bosonic Lagrangian}
\label{bosons}

We closely follow \cite{Forini:2015mca} in order to compute the action for the bosonic fluctuations, see also \cite{Callan:1989nz,Drukker:2000ep}. 
First, we notice that a second order expansion of the $B$-field \eqref{BC-fields} around the classical configuration \eqref{classicalSol} gives
\be
B_2 = \dd \Big[\f{hi \xi\ell_s^2}{2} y_1\dd y_2\Big]\,,
\ee
which is pure gauge, and therefore does not play a role.

In static gauge the quadratic bosonic Lagrangian can be written in terms of fluctuations transverse to the worldsheet, that is 
\be
\label{L-transverse-b}
\mathcal L_{\rm transv}=\sqrt\gamma \left( {\gamma^{ij}} {\rm{D}}_i \zeta^a {\rm{D}}_j \zeta_a -\mathcal M_{ab} \zeta^a \zeta^b\right)\,. 
\ee
The index $a$ is the index on the normal bundle, i.e. transverse directions with respect to the string worldsheet, and so $a=1, \dots, 8$ with a flat metric $\delta_{ab}$. We remind the reader that $i, j$ are the worldsheet curved indices. 
The transverse fluctuations are defined as 
\be
\zeta^{\hat\mu}=\zeta^\mu E^{\hat\mu}_\mu=N^{\hat\mu}_a \zeta^a\,,
\ee
where $\hat\mu$ is the ten-dimensional flat index, $\mu$ is the 10-dimensional curved index, and $N_a^{\hat\mu}$  are 8 orthonormal vector fields orthogonal to the worldsheet. 
The covariant derivative ${\rm{ D}}_i$ is defined as
\be
{\rm{D}}_i \zeta^a =\partial_i \zeta^a -A^a_{b  i} \zeta^b , \qquad \text{with}\qquad A^a_{b  i}= N_b^{\hat\mu}\left(\partial_i N^a_{\hat\mu}-N^a_{\hat\nu}\Omega^{\hat\nu}_{\hat\mu i}\right)\,,
\ee
where $A^a_{b  i}$ is a connection on the normal bundle which contains a term depending on the orthonormal frame $N_a^{\hat\mu}$ and a term with the 10-dimensional spin connection projected down to the classical worldsheet $\Omega^{\hat\nu}_{\hat\mu i}=\Omega^{\hat\nu}_{\hat\mu \mu}\partial_i X^\mu$.
For the classical configuration \eqref{classicalSol} (which essentially does not extend on the compact part of the ten-dimensional target space), the covariant derivative is trivially the derivative along the worldsheet directions:
\be
{\rm{D}}_i= \partial_i \,. 
\ee
The mass term $\mathcal M_{ab}$ in \eqref{L-transverse-b}  is constructed from the ten-dimensional Riemann tensor as well as from the extrinsic curvature
\be
\mathcal M_{ab}= R_{\hat\mu\,\hat\lambda,\,\hat\nu\, \hat\kappa} E^{\hat\lambda}_\mu\, \partial_i X^\mu \, E^{\hat\kappa}_\nu \, \partial^i X^\nu \, N_a^{\hat\mu} N_b^{\hat\nu}+ K_{a, ij} K^{ij}_b\,,
\ee
where $ K^a_{ ij} $ is the orthogonal components of the extrinsic curvature $K^a_{ ij}= K_{ij}^{\hat\mu} N_{\hat\mu}^a$ induced by the classical solution. 
Again for the solution \eqref{classicalSol} the story is quite simple, because all the components of the extrinsic curvature are zero on-shell 
\be
K^{\mu}_{ij}=0\,, 
\ee
and so the contribution to $\mathcal M_{ab}$ comes only from the projected Riemann tensor. 
Indeed this is exactly what happens for the string solution in AdS$_5$ dual to the 1/2-BPS circular WL discussed in \cite{Forini:2015mca}, we can simply read the orthonormal vectors $N_a^{\hat\mu}$ from there. 
This is not surprising since our induced metric on the worldsheet \eqref{classicalMetric} differ by a conformal factor from the induced metric for the circular string  in AdS$_5$, and the classical equations of motion are the same. 
The mass term $\mathcal M_{ab}$ is then given by a diagonal matrix $\mathcal M_{ab}=M_{a} \delta_{ab}$ where 
\be
\begin{split}
M_a &=  R_\gamma- {3\tanh\sigma \over 16\ell_s^2 \xi\cosh^2\sigma} \,, \qquad a=1,2,3,4\,, 
\\ 
M_a &= \frac 14 R_\gamma \,,  \qquad a=5, 6\,,
\\ 
M_a &= \frac 14 R_\gamma+\frac{ h^2}{4 \ell_s^2 \xi} \tanh ^3\sigma\,, \qquad a=7, 8\,,
\end{split}
\ee
where $R_\gamma= -2\e^{-2\rho}\partial_\sigma^2\rho$ is the worldsheet Ricci scalar defined in \eqref{ricciscalar}.

Looking at the ten-dimensional metric \eqref{metricsigma}, the first four directions correspond to the fluctuations along the directions transverse to the equator (parametrized by $\tau$) in the 5-sphere $\sphere^5$, which we denote by ${\bf x} =(x_1,x_2,x_3,x_4)$. 
The fluctuations labelled by $a=5,6$ are local coordinates on ${S}^2$, and we denote them by ${\bf y} = (y_1,y_2)$. 
Finally,  on the classical solution \eqref{classicalSol}, the metric \eqref{metricsigma} is degenerate along the $\phi$ direction, hence we can introduce a two-vector $\bf z=\theta \hat n$ such that ${\bf z} \cdot {\bf z}=\theta^2$ and $\dif{\bf z} \cdot \dif {\bf z} = \theta^2\dif \phi^2+\dif\theta^2$. 
We use then $\bf z$ as local coordinates in this new $\R^2$, which corresponds to $a=7,8$. 
Notice that it's here that the $h$ term enters. 

The final transverse bosonic action \eqref{L-transverse-b} then reads as 
\be
\mathcal L_{\rm transv}= - \zeta^a \delta^{ij} \partial_i \partial_j \zeta_a - e^{2\rho} \mathcal M_{aa} \zeta^a \zeta^a \,. 
\ee
For later use, it is convenient to directly compute $e^{2\rho} \mathcal M_{aa}$, that is
\be
\label{theEs-app}
\begin{split}
E_x &=-e^{2\rho} M_x=  \partial_\sigma^2 \rho+(\partial_\sigma \rho)^2-1\,,\\
E_y &=-e^{2\rho} M_y= \f12 \partial_\sigma^2 \rho\,,\\
E_z &= -e^{2\rho} M_z= \f{1-3h^2}{2}\partial_\sigma^2 \rho+h^2(\partial_\sigma\rho)^2-h^2\,,
\end{split}
\ee
where we used the identities
\be
\begin{split}
\partial_\sigma^2 \rho &= \f{2}{\sinh^2\sigma}-2\Big(\partial_\sigma \rho+\f{1}{\tanh\sigma}\Big)^2 = -\f12 \e^{2\rho}R_\gamma\,,\\
\f{ 2}{\cosh^2\sigma}&=2+3\partial_\sigma^2\rho-2(\partial_\sigma\rho)^2
\end{split}
\ee
to simplify the above expressions. 
In this way the transverse bosonic Lagrangian becomes
\be
\mathcal L_{\rm transv}= - \zeta^a \delta^{ij} \partial_i \partial_j \zeta_a +E_a  \zeta^a \zeta^a\, \equiv  \zeta^a \tilde{\mathcal{K}}_a\zeta^a\,,
\ee
where $ \tilde{\mathcal{K}}_a$ are the flat bosonic operators in \eqref{op-K-Ktilde}, that is 
\be
 \tilde {\cal K}_a=-\partial_\sigma^2 - \partial_\tau^2 + E_a\,, \qquad \text{with}\qquad a=x, y, z\,. 
\ee 
As explained in the main body, we are usually interested in the determinant of operators where we have stripped off the conformal factor, then it is useful to also define the $\mathcal K_a$ operators, as in \eqref{sk}, that is 
\be
{\cal K}_a= \e^{-2\rho}\tilde {\cal K}_a\,. 
\ee

\subsection{Fermionic Lagrangian}
\label{Fermions}

We now turn to the evaluation of the fermionic Lagrange density ${\cal L}_F$.
Our starting point is the quadratic fermionic action \eqref{FermionicAction} which we rewrite here for convenience, that is
\be
\label{FermionicAction-app}
S_\text{fermions} = -\frac{1}{2\pi \ell_s^2}\int\Big\{ i\bar\theta P^{ij} \Gamma_i D_j \theta - \frac{i}{8}\bar\theta P^{ij} \Gamma_{11}\Gamma_i^{\p{i}\mu\nu}H_{j\mu\nu} \theta+\frac{i}{8} \e^{\Phi}\bar\theta P^{ij}\Gamma_i(-\Gamma_{11}\slashed{F}_2+\slashed{F}_4)\Gamma_j\theta\Big\}\,,
\ee
where $\Gamma_\mu$ are the ten-dimensional gamma matrices, $i,j=\sigma,\tau$ are the curved  worldsheet indices, and the projector $P^{ij}$ is defined as 
\begin{equation}
P^{ij} = \sqrt{\gamma} \gamma^{ij} - i\epsilon^{ij}\Gamma_{11}\,,
\end{equation}
where the matrix $\Gamma_{11} $ in our convention reads as 
\begin{equation}
\Gamma_{11} = i\Gamma_{\hat \sigma\hat\tau \hat x_1 \hat x_2\hat x_3\hat x_4 \hat y_1\hat y_2\hat z_1\hat z_2}\,.
\end{equation}
We have written hats on indices to stress that these are tangent target space indices.
We remind the reader that we work in static gauge, which implies $\partial_i X^\mu = \delta_i^\mu$.  

First of all, we notice that there is no contribution from the three-form $H_{3}= d B_{2}$  once projected on the classical string solution \eqref{classicalSol}. 
We can then simplify the action \eqref{FermionicAction-app} as 
\be
S_\text{fermions} = -\frac{1}{2\pi \ell_s^2}\int\Big\{ i\bar\theta P^{ij} \Gamma_i D_j \theta +\frac{i}{8} \e^{\Phi}\bar\theta P^{ij}\Gamma_i(-\Gamma_{11}\slashed{F}_2+\slashed{F}_4)\Gamma_j\theta\Big\}\,. 
\ee
It is useful to define the following projector 
\be
\label{def-P-ferm}
\mathcal P\equiv {\left(\mathbf 1-i \Gamma_{\hat \sigma\hat\tau} \Gamma_{11}\right)\over 2}\,, 
\ee
then we see that 
\be
P^{ij}\Gamma_i\Gamma_j = 2 \sqrt{\gamma}(\mathbf 1-i \Gamma_{\hat \sigma\hat\tau} \Gamma_{11}) = 4 \sqrt{\gamma}{\cal P}\,,\qquad 
\text{and} \qquad 
P^{ij}\Gamma_i =  2 \sqrt{\gamma}\, \Gamma^j {\cal P}\,, 
\ee
where we have used that $\epsilon^{12} = \epsilon^{\sigma\tau} = 1$.

The covariant derivative is defined as
\be
D_j= \partial_j +{1\over 4} \partial_j X^\mu \omega_\mu^{\hat\mu\hat\nu}\Gamma_{\hat\mu\hat\nu}\,,
\ee
where $\hat\mu, \hat\nu$ are flat ten-dimensional indices. 
The only non-zero component of the spin connection along the directions tangent to the worldsheet  is
\be
{\omega_\tau}^{\hat\tau\hat\sigma} = -\f{1}{\tanh\sigma} - \f{1}{\sinh 2\sigma} = \partial_\sigma\rho \,.
\ee
Using this and the fact that the projector $\mathcal P$ commutes with $\Gamma^j$, leads to the following simple expression for the kinetic term in \eqref{FermionicAction-app}:
\be\label{kin-op-ferm}
\slashed{D} = \e^{-\rho}\left(\Gamma_{\hat\sigma}\partial_\sigma + \Gamma_{\hat\tau}\partial_\tau +\f1{2}\partial_\sigma\rho\Gamma_{\hat\sigma}\right)\,.
\ee

The two- and four-form are particularly simple since $H_3 = 0$, and to leading order we have 
\be\label{f2f4}
\begin{split}
F_2  &= \dif C_1= \f{h N\pi i  \ell_s}{\xi}\tanh^2\sigma\,\vol_{\R_z^2}\,,\\
F_4  &= \dif C_3=-3N \pi  \ell_s^3\,\vol_{\R_z^2}\w \vol_{\R^2_y}\,,
\end{split}
\ee
where the potentials are defined in \eqref{BC-fields}. 
Notice that in principle we could  also have a term $H_3 \wedge C_1$ in $F_4$ (see \eqref{BC-fields-v2}).
However, this would contribute to a higher order, and since here we are only interested in quadratic fluctuations, we can safely neglect it. 
From the expressions \eqref{f2f4}, the flux terms in the fermionic Lagrangian become
\be\label{fluxes-ferm}
\begin{split}
\f14\e^{\Phi}\slashed{F}_2  & = \f{ih   }{4\sqrt{\xi}\ell_s}\tanh^{3/2}\sigma\Gamma_{\hat z_1\hat z_2}\,,
\\
\f14\e^{\Phi}\slashed{F}_4  &= -\f{3}{4\sqrt{\xi}\ell_s}\sqrt{\tanh\sigma} \Gamma_{\hat z_1\hat z_2\hat y_1\hat y_2}\,,
\end{split}
\ee
where we also used the on-shell expression for the dilaton \eqref{dilaton}. 

We are ready to collect all the terms. 
The projection operator ${\cal P}$ commutes with the kinetic operator \eqref{kin-op-ferm}, the four-flux $\slashed{F}_4 $ in \eqref{fluxes-ferm}, and with the two-form $\Gamma_{11}\slashed{F}_2 $  \eqref{fluxes-ferm}, that is we can rewrite the fermionic action as 
\be
\label{action-ferm-v2-app}
S_\text{fermions} = -\frac{i}{\pi \ell_s^2}\int \e^{2\rho}\, \bar\theta \Big\{  \slashed{D}  +\frac{1}{4} \e^{\Phi}(\Gamma_{11}\slashed{F}_2+\slashed{F}_4)\Big\}{\cal P}\theta\, . 
\ee
From the above Lagrangian, it is natural to choose a gauge to fix the $\kappa$-symmetry as 
\be\label{kappa-sym}
\mathcal P \theta=\theta\,,
\ee
which reduces the spinors from 32 to 16 components. 
For simplicity we define 
\be
\mathcal{D}\equiv i\Big\{  \slashed{D}  +\frac{1}{4} \e^{\Phi}(\Gamma_{11}\slashed{F}_2+\slashed{F}_4)\Big\}\, =i \big(\slashed{D} + \e^{-\rho}X\big)\,,
\ee
where 
\be
X = \f{h i  }{2\cosh\sigma}\Gamma_{11}\Gamma_{\hat z_1\hat z_2} - \f{3}{2\sinh\sigma} \Gamma_{\hat z_1\hat z_2\hat y_1\hat y_2}\,.
\ee
Then the action \eqref{action-ferm-v2-app} is simply given by 
\be
\label{action-ferm-v3-app}
S_\text{fermions} = -\frac{i}{\pi \ell_s^2}\int \e^{2\rho}\, \bar\theta \mathcal{D} \,\mathcal P \theta\, . 
\ee
We choose a Hermitian basis for the ten-dimensional gamma matrices, then we notice that $\slashed{D}^\dagger =- \slashed{D}$, and $X^\dagger = X$ which implies 
\be
\mathcal{D}^\dagger = i\big(\slashed{D} - \e^{-\rho}X\big)\,,
\ee
that is our fermionic operator is not Hermitian. 
For later convenience, we can write the fermionic operator as
\be\label{tenDfermionop}
\mathcal{D} = \e^{-3\rho/2}\big(i\slashed{\partial} + i X\big)\e^{\rho/2}\,.
\ee
This operator acts on the entire 16 component fermion on the worldsheet (after fixing the $\kappa$-symmetry). 
Our aim is to obtain a two-dimensional fermionic operator. To this end, we can make this manifest by defining $\pm$ eigenspaces of the operators
\be
i\Gamma_{\hat z_1\hat z_2}\,,\quad i\Gamma_{\hat y_1\hat y_2}\,,
\ee
each of the four eigenspace is four-dimensional accounting for all fermionic degrees of freedom. 
In practice we introduce the projectors
\be\label{projectors-s1s2}
\mathcal P_{\pm, \hat z}={\mathbf 1\pm i\Gamma_{\hat z_1\hat z_2}\over 2}\,, \qquad 
\mathcal P_{\pm, \hat y}={\mathbf 1\pm i\Gamma_{\hat y_1\hat y_2}\over 2}\,,
\ee
whose eigenvalues are $\pm 1$. 
We label the eigenspaces by $s_1$ and $s_2$ respectively. 
These should be split further into a pair of two-dimensional fermions. 
We do this by identify the ten-dimensional $\sigma$ and $\tau$ Gamma matrices with Pauli matrices and define proper two-dimensional operators that act on the fermions, that is 
we choose the following representation for the $\Gamma$-matrices
\be
\Gamma_{\hat\sigma}= \tau_1  \otimes {\mathbf 1}_{16} \,,\qquad \Gamma_{\hat\tau} = -\tau_2 \otimes  \mathbf{1}_{16}\,,
\qquad  \Gamma_{\hat p} =\tau_3 \otimes \gamma_{\hat p}\,, \qquad \hat p\neq \hat \sigma, \hat \tau\,,
\ee
where $ \gamma_{\hat p}$ are $16\times 16$ Dirac matrices along the 8 transverse Euclidean directions. 
Combining the $\kappa$-symmetry gauge choice \eqref{kappa-sym} with our representation for the $\Gamma$-matrices results in
\be
\label{gamma11-2d}
\Gamma_{11}\theta= -(\tau_3 \otimes {\mathbf 1}_{16})\theta\,.
\ee
Applying the decomposition \eqref{projectors-s1s2}-\eqref{gamma11-2d} to the operator \eqref{tenDfermionop} leads to the two-dimensional operators 
\be
\label{op-D-s}
\begin{split}
\mathcal{D}(s_1,s_2,s_3)  & = \e^{-3\rho/2}\tilde{\mathcal{D}}(s_1,s_2,s_3)\e^{\rho/2}\,,\\
\hskip 0.7 cm
 \tilde{\mathcal{D}}(s_1,s_2,s_3) &=i\slashed{\partial} + \tau_3 a_{s_1} + v_{s_1s_2}
 \end{split}
\ee
where
\be
a_{s_1} = -i\f{h s_1  }{2\cosh\sigma}\,,\quad v_{s_1s_2} = \f{3 i s_1 s_2}{2\sinh\sigma} \,,
\ee
and each of the 8 two-dimensional fermion is labeled by $s_{1,2,3} = \pm1$.
It is clear that the operators in \eqref{op-D-s} do not depend on the third eigenvalue $s_3$. 
With a view to the computation of functional determinants, this implies that 
\be
\prod_{s_1, s_2,  s_3=\pm 1}  \det \mathcal D({s_1, s_2,  s_3})= 
\prod_{s_1, s_2 =\pm 1} \left(\det \mathcal D({s_1, s_2, 1})\right)^2 \,. 
\ee
Furthermore, denoting the time reflection operator by $\mathbf{T}$, we see that
\be
\mathbf{T}\tau_1\mathcal{D}(s_1,s_2,s_3)\tau_1 = \mathcal{D}(-s_1,-s_2,s_3)\,,
\ee
which allows us to reduce to two determinants. However, there is still a residual symmetry given by
\be
\tau_3\mathcal{D}(s_1,s_2,s_3)\tau_3 = -\mathcal{D}(-s_1,s_2,s_3)\,,
\ee
which, combined with the time reflection, gives us 
\be
\prod_{s_1, s_2,  s_3=\pm 1}  \det \mathcal D({s_1, s_2,  s_3})=  (\det \mathcal D({-1, -1,  1}))^8\,.
\ee
This means that we are left only with one two-dimensional fermionic operator (with a multiplicity 8):
\be
{\mathcal D}\equiv \mathcal D({-1, -1,  1})\,,
\ee
which is the fermionic operator in \eqref{op-D-Dtilde}-\eqref{a-v-ferm}.

\section{Phase shifts for generic values of $h$}
\label{app:phaseshift}


In this section we collect the results for the phase shifts corresponding to the bosonic and fermionic operators for different values of the parameter $h$.
We recall that $h$ was initially defined in \eqref{metricsigma} and \eqref{dilaton-10}. 
As a consequence, $h$ appears in the potential $E_z(\s)$ for the bosonic fluctuations along the $\bf z$-directions, cf. equation \eqref{theEs}, and in the term $a(\s) $ \eqref{a-v-ferm} in the fermionic operator $\tilde{\mathcal D}$ \eqref{op-D-Dtilde}.

\subsection{Phase shifts for bosonic operators}
\label{app:res-various-h}

The phase shifts for the bosonic operators can be found analytically for any value of the parameter $h$. Recall that we are dealing with Hermitian operators, and so the two linear independent solutions are complex conjugated, i.e. $\eta_{a}(p;\sigma)$ and $\bar \eta_a(p;\sigma)$, which implies for the phase shifts
\be
\delta_a(i p,p)= \delta_a(-i p,p)\,, \qquad a=x,y,z\,. 
\ee
We can simply repeat all the steps discussed in section \ref{phaseshiftmethod} in a straightforward way. 
The explicit form of the bosonic basis function is then
\be
\begin{split}
\eta_{x}(p;\sigma) &= \big(2\sinh \sigma\big)^{i p}(\coth\sigma)^{1/2}\, \HyperGeo\Big(-\tfrac{1+i p}{2},\tfrac{3-ip}{2};1-ip;-\sinh^{-2}\sigma\Big)\,, \\
\eta_{y}(p;\sigma) &= \big(2\sinh \sigma\big)^{i p}(\coth\sigma)^{1/2}\, \HyperGeo\Big(-\tfrac{i p}{2},\tfrac{2-ip}{2};1-ip;-\sinh^{-2}\sigma\Big)\,, \\
\eta_{z}(p,h;\sigma) &= \big(2\sinh \sigma\big)^{i p}(\coth\sigma)^{1/2-h}\, \HyperGeo\Big(-\tfrac{h-2+i p}{2},-\tfrac{h+ip}{2};1-ip;-\sinh^{-2}\sigma\Big)\,,
\end{split}
\ee
for the three operators $\tilde{\cal K}_x$, $\tilde{\cal K}_y$, and $\tilde{\cal K}_z$ \eqref{op-K-Ktilde}. 
Thus, the phase shifts for the three bosonic operators for arbitrary $h$ are given by
\be
\begin{split}
\delta_x(p) &= \text{Arg}\, \Big[{2^{-ip}\Gamma(\tfrac{3-ip}{2})^2\Gamma(1+ip)}\Big]\,,\\
\delta_y(p) &= \text{Arg}\, \Big[\Gamma(1-\tfrac{ip}{2})\Gamma(\tfrac{1+ip}{2})\Big]\,,\\
\delta_z(p) &= \text{Arg}\, \Big[{2^{-ip}\Gamma(\tfrac{2-h-ip}{2})\Gamma(\tfrac{2+h-ip}{2})\Gamma(1+ip)}\Big]\,,
\end{split}
\ee
where again the dependence on $h$ of the phase shift is only along the $\bf z$-directions.

\subsection{Phase shifts for fermionic operators}
\label{app:res-various-h-ferm}

Concerning the fermionic operators, we are able to analytically obtain the phase shifts for $h=0$ and $h=3$. 
The reason is due to the fact that the differential equations \eqref{fermioniceq} drastically simplify for these special values of $h$ and reduce to hypergeometric differential equations.
By looking at the explicit expression of the fermionic operator \eqref{a-v-ferm}, we see that for $h=0$ the term $a(\s)$ vanishes, and the operator $\tilde{\mathcal D}$ becomes degenerate, while for $h=3$ the terms $a(\s)$ and $v(\s)$ now enter with the same coefficient. 

The steps to analytically  find the phase shifts are the same as performed numerically, and they are described in section \ref{phaseshiftmethod}.
Here we only  present the results for $h=0$, that is
\be
\label{delta-ferm-h0}
(4\delta_{+}+ 4\delta_{-}) = 8\text{Arg}\, \Big[\Gamma(\tfrac12+i p)\Gamma(2-ip)\Big]\,, 
\ee
and for $h=3$, namely
\be
\label{delta-ferm-h3}
(4\delta_{+}+ 4\delta_{-}) = -4\arctan\f{2p}{5} - 4 \arctan\f{2p}{3}-4\arctan 2p\,.
\ee

\begin{figure}[h]
\centering
\includegraphics[width=0.6\textwidth]{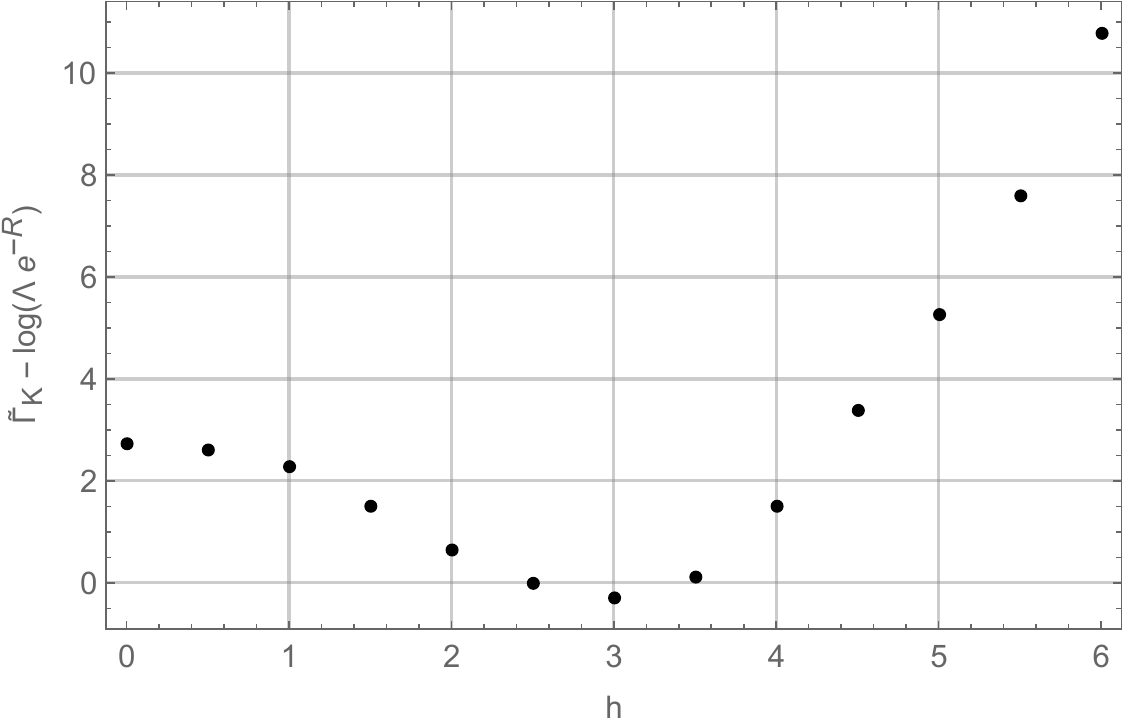}
\caption{\label{Gammaofh} The regularized one-loop effective function $\tilde\Gamma_{\mathbb K}(R)-\log(\Lambda e^{-R})$ \eqref{phaseshiftformula} for different values of $h$, integers and half-integers. }
\end{figure}

\vskip 1 cm
It is natural to ask ourselves how the parameter $h$ affects the one-loop effective action. 
We remind the reader that this parameter is not visible on the matrix model side since it breaks supersymmetry. 
However, it clearly influences the one-loop string partition function $e^{-\tilde\Gamma_{\mathbb K}(R)}$  \eqref{phaseshiftformula}, being part of the geometrical background \eqref{metricsigma}-\eqref{dilaton-10}. For this reason we have computed the finite part of $\tilde\Gamma_{\mathbb K}(R)$ for different values of $h$ as summarised in figure \ref{Gammaofh}.


\section{Phase shifts for circular string in AdS$_4$}
\label{app:Ads}

In this appendix, using the phase shift method,  we compute the functional determinants  for strings in AdS$_n$ ($n=5,4,3$) dual to 1/2-BPS circular Wilson loops. 
The ultimate goal is to find the string one-loop partition function $e^{-\tilde\Gamma_{\mathbb K}(R)}$ for the case $n=4$ in the same regularization scheme adopted throughout this manuscript. This will allows us to safely remove the regulators in \eqref{logZD4result} (and thus safely compare  the corresponding WL expectation values).  
The following calculations have been also performed in~\cite{Cagnazzo:2017sny,Medina-Rincon:2018wjs,Medina-Rincon:2019bcc} with the same method. 

In the language we have been using, we now have the following bosonic operators 
\be
\begin{split}
& \tilde{\mathcal{K}}_{a, {\rm AdS}} = -\partial_\sigma^2-\partial_\tau^2+ E_a(\sigma)\,, \qquad  a=1, 2\,,\\
& E_1  = \f{2}{\sinh^2\sigma}\,,\qquad n-2\text{ times,} \qquad a=1\,,\\
& E_2   = 0\,,\qquad 10-n\text{ times,} \qquad a=2\,, \\
\end{split}
\ee
and  fermionic operators
\be
\begin{split}
&\tilde{\mathcal{D}}_{a, {\rm AdS}}  =i\slashed{\partial} + \tau_3 a_a + v_a, \, \qquad  a=1, 2\,,\\
& a_1 = \f{1}{\sinh\sigma}\,,\qquad v_1=0\,,\qquad  2n-2\text{ times,}  \qquad a=1\,,\\
& a_2 = 0\,,\qquad v_2=0\,,\qquad  10-2n\text{ times,} \qquad a=2\,.  \\
\end{split}
\ee
Notice that all the operators are Hermitian. 
We can now compute the logarithm of the one-loop partition function for the flat operators $\tilde{\mathcal{D}}_{a, {\rm AdS}} $ and $\tilde{\mathcal{K}}_{a, {\rm AdS}}$. 
In particular, for the case $n=4$, the logarithmic of the string partition function we are interested in is given by
\be
\label{Gamma-tilde-AdS4}
\tilde\Gamma_{\mathbb K, {\rm AdS}}(R)= \f12 \log \f{\det^2{\tilde{\mathcal{K}}}_{1, {\rm AdS}} \det^6{\tilde{\mathcal{K}}}_{2, {\rm AdS}}}{\det^6{\tilde{\cal D}}_{1, {\rm AdS}} \det^2{\tilde{\cal D}}_{2, {\rm AdS}}}\,,
\ee
which, in terms of the phase shift, becomes 
\be\label{log-Gamma-AdS-delta}
\tilde\Gamma_{\mathbb K,{\rm AdS}}(R)= -\int_0^\infty \dd p\, \Big[\coth (\pi  p)\delta_B-\tanh (\pi  p) \delta_F\Big] - R\,,
\ee
where we have denoted the sum of the bosonic and fermionic phase shifts by $\delta_B$ and $\delta_F$ respectively. 
These are straightforward to compute, and for general $n$, we find
\be
\delta_B =- (n-2)\,\text{arctan}\,p\,,\qquad \delta_F = -2(n-1)\,\text{arctan}\,2p\,, \qquad n=5, 4, 3\,. 
\ee
We note that the phase shifts are only defined up to factors of $\pi$.
The logarithm of the one-loop partition function \eqref{log-Gamma-AdS-delta} is now UV divergent in two ways. Indeed, for large $p$, we have 
\be
\lim_{p\to\infty}(\delta_B - \delta_F) = \f{n\pi}{2} - \f{1}{p}\,.
\ee
The first term produces linear divergence in the partition function, while the second term produces the expected $\log \Lambda$ divergence. 

We now focus on the AdS$_4$ case with $n=4$ in the above equation. By shifting $\delta_F$ by a factor of $2\pi$ we can eliminate the spurious linear divergence 
\be
\delta_F \mapsto\delta_F + 2\pi\,.
\ee 
Thus, for $n=4$ the one-loop partition function \eqref{log-Gamma-AdS-delta} can now be written as 
\be
\begin{split}
\Gamma_{\text{AdS}_4}  &= -\int_0^\infty \dd p\Big[-2 \arctan p \coth(\pi p) -\left(2\pi-6 \arctan 2p\right) \tanh(\pi p) \Big] -R\,\\ 
 &= 2\log\pi+\log(\Lambda \e^{-R})\,.
 \end{split}
\ee
where in the last step we  analytically evaluated the integral  using the results in \cite{Medina-Rincon:2018wjs}\,.

\newpage
\bibliography{refs}
\bibliographystyle{JHEP}

\end{document}